\documentclass[fleqn,10pt]{wlscirep}
\usepackage[utf8]{inputenc}
\usepackage[T1]{fontenc}
\usepackage{lineno}
\usepackage{fancyhdr}

\title{Dust attenuation evolution in $z \sim 2-12$ galaxies observed by JWST \thanks{This is the accepted manuscript version of the article published in \textit{Nature Astronomy}: Markov, V. et al. (2024), Nat. Astron. \href{https://doi.org/10.1038/s41550-024-02426-1}{https://doi.org/10.1038/s41550-024-02426-1}. The final version is available at: \href{https://www.nature.com/articles/s41550-024-02426-1}{nature.com/articles/s41550-024-02426-1}}}
\author[1,2,$^\dagger$]{Vladan Markov}
\author[1,$^\dagger$]{Simona Gallerani}
\author[1,$^\dagger$]{Andrea Ferrara}
\author[1]{Andrea Pallottini}
\author[1]{Eleonora Parlanti}
\author[1]{Fabio Di Mascia}
\author[2]{Laura Sommovigo}
\author[1]{Mahsa Kohandel}
\affil[1]{Scuola Normale Superiore, Piazza dei Cavalieri 7, 56126 Pisa, Italy}
\affil[2]{Faculty of Mathematics and Physics, University of Ljubljana, Jadranska ulica 19, SI-1000 Ljubljana, Slovenia}
\affil[3]{Center for Computational Astrophysics, Flatiron Institute, 162 5th Avenue, New York, NY 10010, USA}

\affil[$^\dagger$]{vladan.markov@fmf.uni-lj.si; simona.gallerani@sns.it; andrea.ferrara@sns.it}


\begin{abstract}
\textbf{{A sizable fraction of the heavy elements synthesized by stars in galaxies condenses into sub-micron-sized solid-state particles, known as dust grains. Dust produces a wavelength-dependent attenuation, $A_\lambda$, of the galaxy emission, thereby significantly altering its observed properties. Locally, $A_\lambda$ is in general the sum of a power-law and a UV feature ('bump') produced by small, carbon-based grains. However, scant information exists regarding its evolution across cosmic time. Here, leveraging data from 173 galaxies observed by the James Webb Space Telescope in the redshift range $z=2-12$, we report the detection of the UV bump in a $z \sim 7.55$ galaxy (when the Universe was only $\sim 700$ Myr old), and show that the power-law slope and the bump strength decrease towards high redshifts. We propose that the flat $A_\lambda$ shape at early epochs is produced by large grains newly formed in supernova ejecta, which act as the main dust factories at such early epochs. Importantly, these grains have undergone minimal reprocessing in the interstellar medium due to the limited available cosmic time. This discovery offers crucial insights into the redshift-dependent evolution of dust attenuation properties, shedding light on the role of supernovae-driven dust production,  grain size distribution at early cosmic times, and the processes driving dust evolution at later epochs.} }

\end{abstract}

\begin{document}

\flushbottom
\maketitle

\thispagestyle{empty}

\section*{Introduction} \label{intro}


Dust attenuation refers to the absorption and scattering of photons along the line of sight (LOS), due to the intervening interstellar medium (ISM) dust.  The shape of the dust attenuation curve, $A_\lambda$, in galaxies (i.e. the slope of the curve and the strength of the characteristic UV bump at $\sim 2175$ \r{A}) is determined by both the intrinsic dust properties (mass, grain size distribution, chemical composition), and the spatial distribution of dust with respect to stars. In the case of simple dust-to-star geometry, e.g., a point-like background source with a uniform dust screen, the loss of light along the LOS is instead referred to as "dust extinction", which is governed solely by the intrinsic dust properties\cite{2013seg..book..419C, 2020ARA&A..58..529S}. 

The slope of the attenuation curve is determined by the grain size distribution and chemical composition\cite{1977ApJ...217..425M, 1984ApJ...285...89D, 2001ApJ...548..296W}. In particular, a shift towards smaller (larger) grains results in steeper (shallower) shapes\cite{2001ApJ...548..296W}. Likewise, the rise of the characteristic UV bump feature at $2175$ \r{A} is typically associated with the existence of small carbonaceous dust grains\cite{1965ApJ...142.1683S, 1984ApJ...285...89D}, including polycyclic aromatic hydrocarbons (PAHs)\cite{1990A&A...237..215D, 1992A&A...259..614S}. The V-band attenuation, $A_V$, which serves as a proxy of the galaxy dust content along the LOS, has been identified as one of the primary drivers of the shape of the dust attenuation law\cite{2013MNRAS.432.2061C, 2020ApJ...888..108B}. Theoretical studies suggest that this correlation may arise from radiative transfer effects dependent on the amount and spatial distribution of dust relative to different stellar populations \cite{2018ApJ...869...70N, 2020MNRAS.491.3937T}. For instance, the increasing complexity of the dust-to-star geometry flattens the slope of the dust attenuation curve and reduces the prominence of the UV bump, regardless of the initial dust extinction curve\cite{1999ApJS..123..437F, 2018ApJ...869...70N}. 

The dust attenuation/extinction laws of high-$z$ sources remain largely unknown. Only a handful of early star-forming galaxies\cite{2022A&A...663A..50B, 2023A&A...679A..12M, 2023Natur.621..267W}, quasars\cite{2010A&A...523A..85G, 2021MNRAS.506.3946D}, and gamma-ray burst (GRB) afterglows\cite{2011A&A...532A..45S, 2018A&A...609A..62B} have constrained dust curves. Dust laws at high-$z$ are expected to deviate from the well-established empirical dust curves of the local galaxies, such as the  "Calzetti"\cite{2000ApJ...533..682C} attenuation curve for local starbursts, and the Small Magellanic Cloud (SMC)\cite{2003ApJ...594..279G} and the Milky Way (MW) extinction curves\cite{1989ApJ...345..245C}. This redshift ($z$) evolution may arise from several factors such as distinct dust formation processes\cite{2017ApJ...851..152B, 2020A&A...637A..32B}, and dust reprocessing mechanisms\cite{1978ppim.book.....S, 1980ApJ...239..193D} occurring in the more extreme ISM conditions\cite{2019MNRAS.489....1F, 2019MNRAS.487.1689P, 2022MNRAS.513.5621P, 2020MNRAS.495L..22V, 2022A&A...663A.172M} of early galaxies. Additionally, galaxies at intermediate\cite{2009ApJ...706.1364F, 2022MNRAS.516.3532M} and high-$z$\cite{2023MNRAS.522.3986F, 2023A&A...677L...4P} tend to be less dusty and show a clumpier structure and irregular distribution of dust and gas with respect to stars, compared to nearby sources\cite{2018MNRAS.477..552B, 2018MNRAS.478.1170C,  2020A&A...641A..22M, 2022MNRAS.517.5930S}.

In this Letter, we apply our customized version of the \texttt{BAGPIPES} code\cite{2018MNRAS.480.4379C, 2023A&A...679A..12M} to a large sample of star-forming galaxies at $z \sim 2-12$, from the publicly available {\it James Webb} Space Telescope (JWST) spectroscopic observations, to investigate evolutionary trends of dust attenuation laws and their implications on the intrinsic dust properties and dust-to-star geometry.

\section*{Redshift evolution of dust attenuation law}\label{Results}

We employ our customized version of \texttt{BAGPIPES} code\cite{2018MNRAS.480.4379C, 2023A&A...679A..12M} to fit the individual spectra of a sample of 173 galaxies at $z \sim 2-12$ observed by JWST. The code generates the best-fit model to the NIRSpec spectra and provides the best-fit parameters of the SED model from the medians of the posterior distribution. Outputs of the SED fitting process include fundamental physical characteristics of galaxies (stellar mass ($M*$), star formation rate (SFR), metallicity ($Z$), mass-weighted stellar age ( ${\langle a \rangle}_*^{\rm{m}}$), ionization parameter $U$, and $V$-band dust attenuation ($A_V$)) along with the properties of the dust attenuation model ($c_1 - c_4$; see the SED fitting section for details). We utilize the $c_1 - c_4$ parameters to construct the dust attenuation curve for each galaxy. The inferred attenuation curves of our sample exhibit a wide diversity of slopes and UV bump strengths (see Fig. \ref{all_curves} and the Galaxy Sample section). 
To explore potential trends in dust attenuation evolution, we split our sample in redshift bins by adopting equidistant cosmic time in each bin ($\Delta t \sim 0.5 \ \rm{Gyr}$). The number of sources per bin varies in the 13-66 range.

Fig. \ref{all_curves_bin} depicts the median dust curves with the corresponding 1$\sigma$ dispersion for each subsample. Although the associated uncertainties are relatively large, a noticeable trend emerges in the evolution of the dust attenuation curves with redshift. Specifically, with increasing redshift, the median slope flattens and the strength of the UV bump decreases.

 To provide a more quantitative analysis of these results, we characterize the shape of the dust attenuation curve with the $S$ and $B$ parameters, which represent proxies for the UV-optical slope and the 2175 \r{A} bump strength, respectively\cite{2020ARA&A..58..529S}. The slope is defined by the ratio of attenuation at $1500$ \r{A} and the one in the $V$-band, $S  = A_{1500}/A_V$. The UV bump strength is defined as $B = A_{\rm{bump}}/A_{2175}$, where $A_{\rm{bump}}$ is the additional attenuation above the baseline at 2175 \r{A}, and $A_{2175}$ is the total attenuation at 2175 \r{A} (see the Attenuation curve parametrization section). The trends that we see in Fig.  \ref{all_curves_bin} remain apparent when we plot the UV-optical dust curve slope ($S$) and UV bump strength ($B$) separately, as functions of redshift (Figs. \ref{z_S}a and \ref{z_S}b, respectively).

\subsection*{Redshift evolution of the slope} \label{slope_z}

Fig. \ref{z_S}a illustrates the redshift evolution of the UV-optical slope ($S$), for each source of our full JWST sample, and median values for subsamples grouped by redshift (as in Fig. \ref{all_curves_bin}). The median slope of $z \sim 2-3.5$ galaxies approaches the SMC slope\cite{2003ApJ...594..279G}. However, it gradually flattens with increasing redshift ($z \sim 3.5-5.5$) and at $z > 5.5$ for galaxies is flatter than the MW slope\cite{1989ApJ...345..245C}.
The median slopes of galaxies at intermediate-$z$ are slightly steeper (although they are consistent within the 1$\sigma$ uncertainties) with respect to most of the inferred slopes of various samples of  intermediate-$z$ galaxies from the literature. The discrepancy can be attributed to the fact that our overall JWST sample has a median $A_V$ of $\sim 0.38$. Conversely, the values of $A_V$ for most galaxy samples in the literature are somewhat higher, $A_V \sim  0.9$\cite{2011A&A...533A..93B}, ${A_V} \gtrsim  0.5$ \cite{2012A&A...545A.141B, 2013ApJ...775L..16K, 2015ApJ...800..108S, 2015ApJ...806..259R}, $A_V \sim  0.2-1.1$\cite{2018MNRAS.475.2363T}, and $A_V \sim  0.3-0.9$\cite{2020ApJ...899..117S}. Therefore, this difference in slopes is a consequence of a well-known trend wherein galaxies with low $A_V$ tend to have steeper UV-optical slopes\cite{2013MNRAS.432.2061C, 2018ApJ...869...70N, 2022A&A...663A..50B}. If we instead restrict our JWST sample to more attenuated galaxies (with $A_V > 0.38$), our results are more aligned with the slopes inferred for known samples of $z \sim 1-3$ galaxies. Additional potential origin of this discrepancy is discussed in the Observational and model uncertainties section. 

Fig. \ref{z_S}a also depicts the best fit on the entire sample whereas Fig. \ref{z_S_Av} shows the best fit on subsets of sources with $A_V \sim 0.1-0.38$ and $A_V \geq 0.38$. We see that the slope of the attenuation curve (at fixed redshift) flattens with increasing $A_V$ (despite significant uncertainties), as expected. This effect is more evident at lower redshifts ($z \lesssim 4$), whereas at $z \gtrsim 6$ the dust attenuation curves remain shallow, and are largely unaffected by variations in $A_V$. Furthermore, the analysis of $A_{\lambda}(z)$ in different $A_V$ ranges shows that the flattening trend towards high-$z$ persists irrespective of the $A_V$ values considered.  This implies that $A_V$ is not the driver of the redshift evolution of the slope. Additionally, we examine if the flattening trend stems from an underlying dependence on other galaxy properties (e.g., $M*$, SFR, $Z$, ${\langle a \rangle}_*^{\rm{m}}$, and $U$), but the overall trends remain regardless of the considered parameters (Extended Data Figs. \ref{z_S_logM}, \ref{z_S_age}, \ref{z_S_SFR}, \ref{z_S_Z}, and \ref{z_S_logU}).    

\subsection*{Redshift evolution of the UV bump strength}

Fig. \ref{z_S}b depicts the redshift evolution of the UV bump strength ($B$). Similarly to the observed trends with the slope $S$, $B$  also exhibits a general decrease as redshift increases. The median $B$ value reaches levels comparable to $\sim 20-25\%$ of the MW bump strength in the $z \sim 2-4$ range. This is comparable with the inferred $B$ values of the intermediate-$z$ galaxies found in the literature\cite{2011A&A...533A..93B, 2013ApJ...775L..16K}, although the uncertainties remain relatively high. To further support this result, we emphasize that if we characterize the UV bump strength using the $c_4$ parameter (Eq. \ref{dust_law}\cite{2008ApJ...685.1046L}), we observe comparable trends in the redshift evolution of the UV bump strength (Extended Data Fig. \ref{z_c4}).

The presence of the UV bump has been confirmed with a significance level of $>3\sigma$ ($B/B^{\rm{err}} > 3$), for 28 sources, within the $z \sim 2-8$ range (marked by red open circles in Fig. \ref{z_S}b). This accounts for $\sim 16\%$ of the full JWST sample. However, $B$ is not well constrained and exhibits high uncertainties due to error propagation. If instead we use the $c_4$ parameter, which is better constrained as an output of the SED fitting procedure, we obtain a securely detected UV bump in  $\sim 21\%$ of the sources (red open circles in Extended Data Fig. \ref{z_c4}). This percentage ($\sim 21\%$) is consistent with previous results \cite{2009A&A...499...69N, 2012A&A...545A.141B}  found in samples of $z \sim 1.0-2.5$ galaxies. The median bump strength of sources with UV bump detection is ${B} \approx 0.7 \times B^{\rm{MW}}$, i.e., ${c_4} \approx 0.75 \times \ {c_4}^{\rm{MW}}$. 

Applying both $B$ and $c_4$ parameterizations, we confirm that the highest redshift galaxy of our JWST sample with a reliable UV bump detection is the 1433\_3989 source at $z \sim 6.14$ (Fig. \ref{spec}a-c). The UV bump strength for 1433\_3989 is measured as $B = 0.43 \pm 0.05$ ($>9\sigma$) or $c_4 = 0.083 \pm 0.014$ ($>6\sigma$), which corresponds to a factor of $\sim 1.2$ or $\sim 1.6$ higher than the MW bump, depending on the considered parametrization. Moreover, the highest redshift source with a confirmed UV bump detection using only the $B$ parameter is the 2750\_449 source at $z \sim 7.55$ (Fig. \ref{spec}d-f). The UV bump strength for 2750\_449 is quantified as $B = 0.34 \pm 0.10$ ($> 3\sigma$) or $c_4 = 0.054 \pm 0.025$ ($> 2\sigma $) (i.e., $\sim 0.95$ or $\sim 1.05$ of the MW bump).

\section*{Discussion} \label{Discussion}

Our results show that the attenuation law evolves with redshift. Specifically, the attenuation slope flattens (from $S>4$ at $z\sim 2$ to $S \sim 2$) at $z> 8$. This trend is found independently of the value of $A_V$. Furthermore, the strength of the UV bump decreases and eventually disappears (from $B\sim 0.1$ at $z\sim 2-4$ to $B\sim 0$ at $z> 5.5$) at early epochs.

We now turn to the interpretation of these findings. In principle, the observed evolutionary trends can be explained by three scenarios: radiative transfer (RT) effects dependent on (1) the amount of attenuation, i.e. $A_V$, or (2) dust distribution relative to stars of various ages, or  (3) changes in the intrinsic dust properties (primarily grain size distribution and chemical composition). Although in general identifying the dominant factor is challenging\cite{2013ARA&A..51..393C}, the quality of the present data allows us to draw solid conclusions. 

It is well established that smaller (larger) $A_V$ values produce steeper (flatter) attenuation curves.  Such correlation has been both predicted\cite{2013MNRAS.432.2061C, 2018ApJ...869...70N, 2020MNRAS.491.3937T} and observed\cite{2016ApJ...827...20S, 2020ARA&A..58..529S}. The physical interpretation is that, in the low-$A_V$ (optically thin) limit, UV photons have a higher probability of being scattered and eventually absorbed, leading to the steepening of the slope\cite{2013MNRAS.432.2061C, 2023MNRAS.519.2475H}. In the  high-$A_V$ (optically thick) limit, the shape of the curve is flattened by both scattering into the LOS and radiation from unobscured OB stars \cite{2018ApJ...869...70N}. 
Indeed, we also find (Fig. \ref{z_S_Av}) that \textit{at fixed redshift} S increases with decreasing $A_V$. However, we do see that the flattening trend towards high-$z$ persists irrespective of  $A_V$. We can then rule out the first scenario. 

Scenario (2) is based on the evidence that complex dust-to-star geometries flatten the slopes and reduce the UV bump strengths of dust attenuation curves\cite{1999ApJS..123..437F, 2000ApJ...539..718C, 2018ApJ...869...70N, 2020MNRAS.491.3937T} as in those conditions UV photons from young stars escape more easily. Although the detailed distribution of dust and gas in galaxies is often not easy to measure, a clumpy ISM structure is commonly reported for galaxies across the entire redshift range $z \sim 2-11.5$ considered here \cite{2018MNRAS.477..552B, 2018MNRAS.478.1170C, 2019MNRAS.489.2792Z, Fujimoto24, Adamo24}. This suggests that the overall complex dust-to-star geometry at $z \sim 2-11.5$ alone may not be sufficient to explain the observed trends in attenuation curves across the considered redshift range. Thus, scenario (2) seems unlikely. 

Scenario (3)  invokes changes in the grain properties. First, we note that our results are consistent with predictions from theoretical studies\cite{2018MNRAS.478.2851M, 2020MNRAS.492.3779H, 2022MNRAS.517.2076M} that have investigated the cosmological evolution of interstellar dust. A recent study\cite{2022MNRAS.517.2076M} finds that, at a fixed metallicity, extinction curves flatten and the UV bump becomes less prominent with redshift. The study concluded that such trends are driven by a shift in the grain size distribution, favouring larger grains at early epochs. What causes such a shift?

The early prevalence of large grains could be attributed to dust production mechanisms in the ejecta of core-collapse Type II supernovae (SNII\cite{2001MNRAS.325..726T, 2007ApJ...666..955N, 2014Natur.511..326G}). These dust sources are predicted to predominantly form large ($ \approx 0.1 \mu$m) grains\cite{2007ApJ...666..955N, 2013EP&S...65..213A, 2014Natur.511..326G, 2018MNRAS.478.2851M}. Grains are then released into the ISM where they undergo shattering processes\cite{2013MNRAS.432..637A, 2018MNRAS.478.2851M, 2020MNRAS.492.3779H, 2022MNRAS.517.2076M} breaking them into smaller units.  However, shattering is a rather slow process, with typical timescales of $\gtrsim 0.5-1 \ \rm{Gyr}$\cite{2013EP&S...65..213A, 2018MNRAS.478.2851M}.  Consequently, the transition from large to smaller grains is quite gradual and is completed only by $z\approx 6$.  Alternatively, small grains might be preferentially destroyed by Coulomb explosions in the presence of strong radiation fields \cite{2020ApJ...892..149T, 2020ApJ...892...84T} as the ones present in early, compact galaxies. 

The redshift evolution of $S$ and $B$ (or $c_4$) we find supports the third scenario. Importantly, the slope remains consistently shallow with a small dispersion at $z \gtrsim 5.5$. This might be interpreted with the existence of a single, dominant dust formation channel (i.e. SNII), in which large grains are formed and only minimally reprocessed in the ISM in the limited cosmic time available\cite{2014Natur.511..326G, 2018ApJ...868...62G}.  The decrease of the bump strength towards high-$z$ also supports this interpretation, as SNII  yields ($\sim$90\% of silicate and $\sim$10\% of carbon dust) are scarce of carbonaceous grains\cite{2020MNRAS.492.3779H}.

At $z \lesssim 5-5.5$, a larger variety of attenuation curves is observed. This is likely due to a combination of several effects, including the transition from SNII to asymptotic giant branch (AGB) stars\cite{1997A&A...326..305W, 2006A&A...447..553F} as the dominant dust sources, dust growth in the ISM\cite{2013EP&S...65..213A, 2019A&A...624L..13L}, more efficient shattering and sputtering\cite{2013MNRAS.432..637A, 2018MNRAS.478.2851M, 2020MNRAS.492.3779H, 2022MNRAS.517.2076M}. Concurrently, the UV bump emerges at $z \lesssim 5$ (around $\sim 1.2 \ \rm{Gyr}$), marking the epoch when dust formation in AGB stars becomes dominant, leading to enhanced production of carbonaceous grains. Finally, a marginal decrease in both $S$ and $B$ at $z \lesssim 3.4$ may be attributed to processes such as coagulation and accretion becoming more prominent at later epochs\cite{2013MNRAS.432..637A, 2020MNRAS.492.3779H}.

In summary, the flattening of dust attenuation curves at high-$z$ suggests that we might be observing galaxies through a curtain of dust newly formed in SNII ejecta, although other less likely explanations cannot be completely excluded.  Progress might come from cosmological hydrodynamic simulations including dust and related RT effects\cite{2018MNRAS.477..552B, 2021MNRAS.506.3946D, 2024arXiv240218515D}. These will make it possible to accurately determine the dust-to-star geometry, incorporate and explore different intrinsic dust properties (grain size distribution and chemical composition), and take into account RT effects. In parallel, insights into the intrinsic dust properties of high-$z$ sources can be gained via observations of point-like sources, such as quasars\cite{2004Natur.431..533M, 2010A&A...523A..85G}, and gamma-ray burst afterglows\cite{2011A&A...532A..45S, 2018A&A...609A..62B}.

\clearpage

\section*{Methods} \label{Methods}

\subsection*{Spectroscopic data} \label{Data}

We use publicly available Near Infrared Spectrograph (NIRSpec) data retrieved from the DAWN JWST Archive (DJA)\cite{DAWN}, an initiative of the Cosmic Dawn Center. DJA includes all publicly available Cycle 1 and 2 spectroscopic and photometric observations from various JWST surveys. The data reduction was performed entirely and uniformly with \texttt{GRIZLI} (Grism redshift and line analysis software for space-based slitless spectroscopy\cite{brammer_2023_8370018}) and \texttt{MSAEXP} (Tool for extracting JWST NIRSpec Micro-Shutter Assembly (MSA) spectra\cite{Brammer_msaexp_NIRSpec_analyis_2022}) Python scripts, by the Cosmic Dawn Center.

Here, we focus on the analysis of NIRSpec prism spectra. The output 1D spectra achieve wavelength-dependent resolution of $R \sim 30-300$, ranging from $\lambda \sim 0.6-5.3 \ \mu m$. This allows us to probe the nebular lines and continuum emission from rest-frame UV to IR. However, due to the limited velocity resolution of $\gtrsim 1000\ \mathrm{km \ s^{-1}}$, we are unable to resolve nebular emission lines. A fraction of the spectral region below the wavelength of the Ly$\alpha$ line is selectively masked due to the potential contamination by the intervening neutral intergalactic medium (IGM).

\subsection*{Sample selection}

Our target selection process closely follows the one outlined in our recent work\cite{2023A&A...679A..12M}. We conduct a visual inspection of all grade 3 NIRSpec prism spectra, opting for galaxies exhibiting a prominent continuum and nebular line emission while excluding sources with strong negative features.  

We select sources at $z>2$ to ensure that we can effectively probe the rest-frame UV-optical wavelength range, allowing us to accurately constrain the shape of the dust attenuation curve and the potential presence of the UV bump. We further clean our sample by selecting spectra with an average channel Signal-to-Noise (S/N) ratio of $\mathrm{S/N} > 3$, in the rest-frame wavelength range of $1925-2425$ \r{A}, corresponding to the peak position of the UV bump ($2175$ \r{A}). Additionally, we conduct a visual examination of the spectral energy distribution (SED) fits for all sources observed with NIRSpec. Any objects for which the SED models cannot yield a reasonable fit within the associated uncertainties are excluded from further consideration.

Finally, we exclude galaxies that exhibit negligible dust attenuation (i.e., with $A_V \sim 0$) from our initial sample. For a galaxy to be considered "dusty", it must meet two criteria: (1) $V-$band dust attenuation of $A_V > 0.1$; (2) $A_V > 0$ with a significance of $> 3\mathrm{\sigma}$ ($A_V/A_V^{\rm{err}} > 3$). 

\subsection*{Galaxy sample}

We ended up with a final sample of 173 "dusty" galaxies. These galaxies fall within a spectroscopic redshift range of $2.0 \lesssim z \lesssim 11.4$, with a median redshift of ${z} \approx 4.5$), with stellar masses of $6.9 \lesssim \log{M_*/M_{\odot}} \lesssim 10.9$, star formation rates $0.1 \lesssim \rm{SFR}/M_{\odot} \ \rm{yr^{-1}} \lesssim 240$ and $V$-band dust attenuation $0.1 \lesssim A_V \lesssim 2.3$ (Supplementary Fig. 1). 

\begin{figure*}
\centering
\includegraphics[width =\hsize]{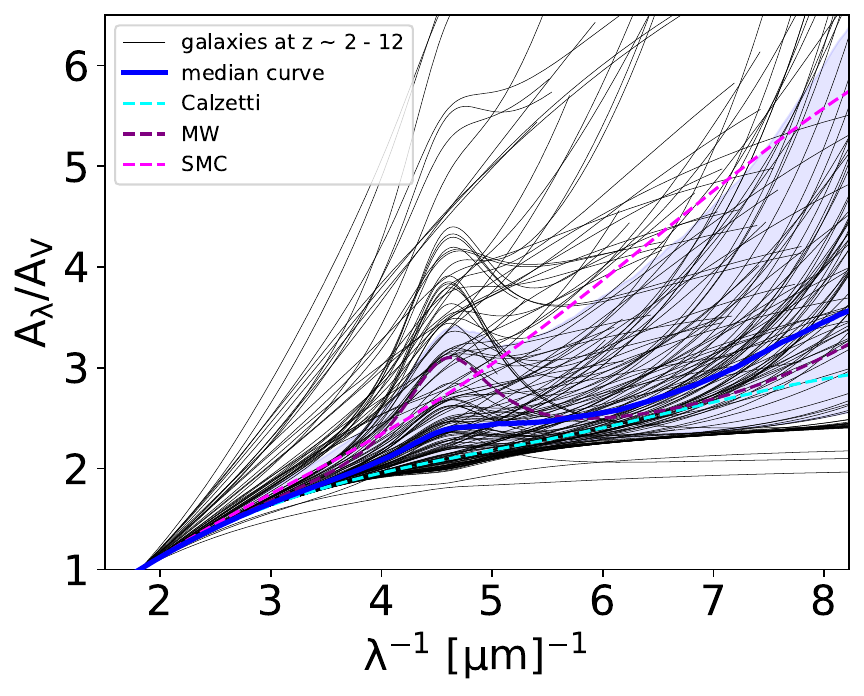}
\caption{{\bf Dust attenuation curves of our full sample of galaxies at $z \sim 2-12$, derived using our SED fitting method.} The median curve of the full sample is depicted as a blue solid line. The corresponding shaded region represents its $1\sigma$ dispersion, obtained with the bootstrapping procedure (see Attenuation curve parameterization section for details). The Calzetti, the MW, and the SMC,  empirical curves are shown as cyan, purple, and magenta dashed lines, respectively.
\label{all_curves}
}
\end{figure*}

\begin{figure*}
\centering
\includegraphics[width=\hsize]{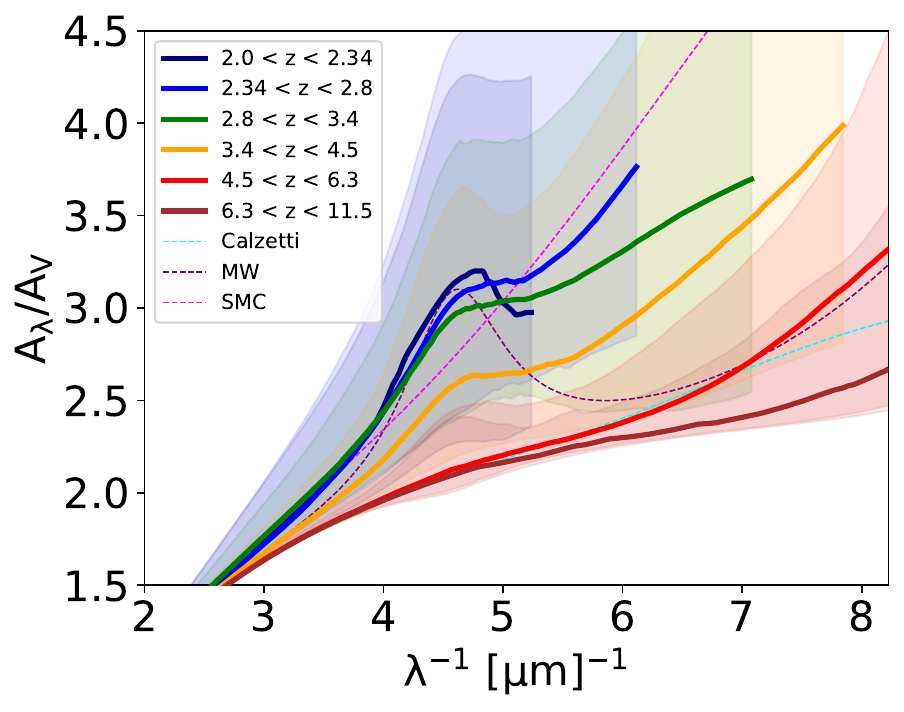}
\caption{{\bf Dust attenuation curves of our full sample of $z \sim 2-12$  galaxies binned by redshift.} The median curves of subsamples are depicted as solid lines. The corresponding shaded regions illustrate their $1\sigma$ dispersion, obtained with the bootstrapping procedure (see Attenuation curve parametrization section). The Calzetti, the MW, and the SMC empirical curves are shown as cyan, purple and magenta dashed lines, respectively.
\label{all_curves_bin}
}
\end{figure*}

Fig. \ref{all_curves} depicts a large diversity of attenuation curves of our full JWST sample. For instance, the slopes range from those shallower than the Calzetti curve to the ones considerably steeper than the SMC curve. The strength of the UV bump varies considerably from curves in which it is completely absent (as in the SMC case) to those surpassing the bump strength observed in the MW. In Fig. \ref{all_curves} we also plot the median attenuation curve along with its 1$\sigma$ uncertainties, computed from the full JWST sample. The median curve and its dispersion are derived using a bootstrapping approach. 

\subsection*{Slit-loss correction}

The flux values at a given wavelength reported in JWST NIRSpec micro-shutter assembly (MSA)-based data may not always coincide with the ones reported in photometric data. The discrepancy between the NIRSpec MSA spectra and the photometric data can arise from several factors such as slit-losses, flux calibration of both spectra and photometry, and spectroscopic and photometric aperture. Slit-loss refers to the reduction in the observed flux in spectroscopic data due to extension and off-centred location of the science target in the NIRSpec spectrograph shutters of finite dimensions ($0.20" \times 0.46"$)\cite{JWST-docs, 2016SPIE.9910E..1OB}. The impact of slit-loss is wavelength-dependent since the full width at half maximum (FWHM) of the point spread function (PSF) widens with increasing wavelength. 

To address the impact of slit-loss and correct for the discrepancy between the NIRSpec spectra and the photometric data, we employ the following approach. First, we utilise the Python tool by DJA\cite{DAWN} to extract photometry from {\it Hubble} Space Telescope (HST), JWST Near-Infrared Camera (NIRCAM), and JWST Near-Infrared Imager and Slitless Spectrograph (NIRISS), using the right ascension, declination and redshift for each target. Next, we overlay the photometric points onto the spectrum for comparison. The discrepancy between the spectroscopic and photometric data is observed in $94\%$ of the sources from our sample. It can reach up to a factor of $6-9$ in the most extreme cases of very extended and/or off-centred sources. In case of a discrepancy, we correct the spectrum by dividing it with one of the throughputs provided by the JWST user documentation\cite{JWST-docs}, until achieving a satisfactory alignment between the spectra and photometry. 

To assess potential systematic uncertainties arising from the above-mentioned slit-loss correction method reliant on visual inspection, we adopt an alternative approach to take into account slit losses. This method involves quantifying the discrepancy between photometric and spectroscopic flux density measurements for each source ($\rm{f_{\lambda}^{phot}/f_{\lambda}^{spec}}$) and performing an uncertainty-weighted linear fit on these quantities. Subsequently, we utilize this linear regression function to correct the spectra affected by slit losses. By applying this reproducible and consistent methodology across our entire galaxy sample, we minimize potential inconsistencies and systematic errors. We find that our results closely mirror the initially reported redshift trends of dust attenuation curves (Figs. \ref{all_curves_bin} and \ref{z_S}), as demonstrated in supplementary Fig. 2. 

Moreover, in Supplementary Fig. 3, we present the redshift trends of the attenuation curve, as well as the $S$ and $B$ parameters derived from SED fitting on spectra without correction for slit losses. In summary, our primary findings regarding the overall redshift trends of the slope ($S$) and the bump ($B$) remain consistent regardless of the application of slit-loss correction or the method employed for it.

\subsection*{FIR observations}

Although a significant portion of our sample sources is expected to exhibit sufficient brightness in the continuum emission within the FIR spectrum to be detectable with submillimeter facilities, given their typical SFR around $\rm{SFR \sim 4 M_{\odot}}$, we delved into the available ALMA data. A considerable fraction of our sources ($>60\%$), remained beyond the observational reach of ALMA due to their positioning in the northern sky, with $\delta > 20^\circ$. The remaining 69 sources of our JWST sample are distributed across the GOODS-South (50), Abell 2744 (17), and RXJ 2129 (2) fields. These sources include 13 (17) sources exhibiting UV bump detections, based on the $B$ ($c_4$) parameter.

Out of these 69 galaxies, ALMA data is available for 62 of them. Specifically, 49 out of 50 sources from GOODS-S were part of the GOODS-ALMA (2015.1.00543.S; PI: D. Elbaz)\cite{2018A&A...620A.152F}, ALMA Hubble Ultra Deep Field (2012.1.00173.S; PI: J. Dunlop)\cite{2017MNRAS.466..861D}, HUDF-JVLA-ALMA (2015.1.00098.S, PI: K. Kohno)\cite{2018PASJ...70..105H}, and GOODS-S (2017.1.00755.S; PI: D. Elbaz)\cite{2022A&A...658A..43G} 1.1-1.3 mm surveys. Additionally, 13 out of 17 Abell 2744 sources were surveyed in MOSAIC-D (2022.1.00073.S; PI: S. Fujimoto)\cite{2023arXiv230907834F} and Abell\_2744 (2018.1.00035.L; PI: K. Kohno)\cite{2022ApJ...932...77S} studies. Finally, ALMA did not observe the two sources within the RXJ 2129 field. Despite the accessibility of ALMA data for these 62 sources, none of them have produced statistically significant detections across the various surveyed datasets.

\subsection*{SED fitting} \label{methodology}

In our recent study\cite{2023A&A...679A..12M}, we presented a customized version of the \texttt{BAGPIPES} SED fitting code\cite{2018MNRAS.480.4379C} that enables us to simultaneously constrain the fundamental physical properties of galaxies (e.g. SFR, $M_*$, $Z$) and the shape of their dust attenuation law. Our tool has been thoroughly tested on synthetic spectra attenuated by empirical dust curves\cite{2023A&A...679A..12M}. In this section, we give a summary of our method. 

\texttt{BAGPIPES} is a Bayesian SED fitting code that generates realistic galaxy spectra using the Stellar Population Synthesis (SPS) models\cite{10.1093/mnras/stw1756}, nebular emission models pre-computed with \texttt{CLOUDY} photoionization code\cite{2017RMxAA..53..385F}, and InterGalactic Medium (IGM) models\cite{2014MNRAS.442.1805I}. The Star Formation History (SFH) can be parameterized with various parametric and flexible SFH models. We employ the non-parametric SFH model with a "continuity" prior as our fiducial SFH model, since it allows greater flexibility in fitting the "true" SFHs and constraining less biased physical properties of galaxies, compared to parametric models\cite{2019ApJ...876....3L, 2022MNRAS.516..975T, 2023A&A...679A..12M}. 

The warm dust emission component within HII regions is included in \texttt{CLOUDY}, whereas the cold dust emission component within the neutral ISM is modelled with the grey body emission. The dust attenuation recipe includes a two-component dust screen model\cite{2000ApJ...539..718C} to account for extra attenuation toward the stellar birth clouds. Dust attenuation can be modelled using standard empirical templates: "Calzetti"\cite{2000ApJ...533..682C} attenuation curve, SMC\cite{2003ApJ...594..279G} and MW extinction curves\cite{1989ApJ...345..245C} and flexible dust models\cite{2000ApJ...539..718C, 2018ApJ...859...11S}. We implement the analytical dust attenuation law\cite{2008ApJ...685.1046L} into \texttt{BAGPIPES} and we use it as our fiducial dust attenuation model. The analytical form of the dust attenuation law, normalised by $A_V$ is:

\begin{equation} \label{dust_law}
\begin{split}   
A_{\lambda}/A_V &= \frac{c_1}{(\lambda/0.08)^{c_2}+(0.08/\lambda)^{c_2} +c_3} \\ 
& +  \frac{233[1-c1/(6.88^{c_2}+0.145^{c_2}+c_3)-c_4/4.60]}{(\lambda/0.046)^2+(0.046/\lambda)^2+90} \\ 
& + \frac{c_4}{(\lambda/0.2175)^2+(0.2175/\lambda)^2-1.95},
\end{split}
\end{equation}%
where $c_1, ...,c_4$ are dimensionless parameters and $\lambda$ is the wavelength in $\mu$m. The three components of the Drude model describe the rise in far ultraviolet (FUV) attenuation, attenuation within the optical and near-infrared (NIR) spectrum, and the $2175$ \r{A} bump, respectively. The main advantage of this flexible dust model is its capability to replicate local empirical dust curves (Calzetti, MW, SMC), along with all the fundamental properties of (synthetic) galaxies\cite{2023A&A...679A..12M}. Additionally, we note that the dust attenuation model flexibility enables the recovery of any potential variations in the dust attenuation laws of early galaxies\cite{2023A&A...679A..12M}. This aspect is significant in the case of high-redshift galaxies since there are no physical reasons to support the notion that dust attenuation curves in early epochs resemble any of the dust laws observed in the local Universe. In this work, we apply our tool to a large sample of high-redshift galaxies, and we showcase the flexibility of the dust model to capture large variations in both the slope of the attenuation curve and the UV bump strength. Nevertheless, it is important to note that the dust attenuation model sets the central wavelength and FWHM of the UV bump at fixed values of $\lambda_0 = 2175$ \r{A} and $\rm{FWHM} \sim 475$ \r{A}, respectively. The central wavelength of the UV bump feature exhibits minimal variation, at least locally\cite{1989IAUS..135..313D}. Hence, fixing the central wavelength at $2175$ \r{A} is a reasonable assumption to make. Nonetheless, several studies of the FWHM of the UV bump at intermediate redshifts ($z \sim 2$) have reported smaller FWHM values, of a factor of $> 2$. This restriction might limit the model's ability to capture variations of these UV bump parameters, which may be observed at higher redshift\cite{2022MNRAS.514.1886S, 2023Natur.621..267W}.  

\texttt{BAGPIPES} is based on a Bayesian approach and the MultiNest nested sampling algorithm\cite{2019OJAp....2E..10F}. \texttt{BAGPIPES} requires to provide the instructions on the SED model, that is, the parameters of the model components (e.g., SFH, nebular emission, and dust attenuation) and their priors. The prior probability distributions for each of the parameters of the model are given in Supplementary Table 1. 
In this work, the selected prior probability densities for model parameters are nearly identical to those of our initial work\cite{2023A&A...679A..12M}. The priors on the the dust attenuation parameters are carefully selected to encompass the parameter range of the dust model fit to Calzetti, SMC, LMC, and MW templates, as well as any potential unconventional attenuation curve variations. However, the allowed prior limits on all parameters have been expanded to encompass a potentially wider parameter space that could be relevant for a larger galaxy sample spanning a significantly broader redshift range of $z\sim 2-12$. For instance, prior limits are set to allow super-solar metallicities (up to $Z \sim 2.5 \ Z_\odot$), high dust content ($A_V \sim 8 $, and prominent UV bump ($c_4 \sim 0.1$). 
The prior limits of the UV bump parameter $c4$ (related to $B$) are set as $-0.005 < c_4 < 0.01$. These limits are designed to accommodate the potential convergence of $c4$ towards zero, a trait observed in certain well-known attenuation/extinction curves lacking a distinct UV bump, such as the SMC and Calzetti curves. As a result, there is a possibility for $c_4$ to occasionally yield slightly negative values during SED fitting (as seen in Extended Data Fig.  \ref{z_c4}). However, it is important to note that such negative values are most probably not indicative of any physical phenomenon but rather a consequence of the constraints imposed by our prior limits.

\subsection*{Statistical tests}

To evaluate the quality of the fit, we initially rely on visual inspection to assess how accurately the SED model captures the features and trends present in the observed spectra.
Additionally, we compute the residuals between the observed and modelled spectra, as illustrated in Figs. \ref{spec}b and \ref{spec}e.

Next, we adopt a frequentist $\rm{\chi^2}$ test approach for assessing the goodness-of-fit and consistency for Bayesian models from the SED fitting procedure\cite{2016A&A...588A..19L}. Specifically, we utilize the chi-square statistics. Here, $\chi^2$ is defined as the difference between the observed and the modelled flux densities normalized by the observed errors:

\begin{equation}
\rm{\chi^2 = \sum_{i=1}^{n} \frac{(f_{\lambda}^{obs} - f_{\lambda}^{model})^2}{\sigma_{obs}^2}}.
\end{equation}

The reduced chi-square statistic $\chi_\nu^2 = \chi^2/\nu$, represents the total chi-square of each source $\chi^2$, normalised by the degrees of freedom, $\nu$. The degrees of freedom are calculated as $\nu = n - k$, where  $n$ denotes the number of data points, while $k$ signifies the number of model parameters\cite{2023A&A...679A..12M} ($k = 16$ for our fiducial model; Supplementary Table 1).
Our analysis reveals that the distribution of $\chi_\nu^2$  is in the range of $\chi_\nu^2 = 4.1_{1.5}^{6.3}$. Notably, three sources exhibit $\chi_\nu^2 >100$, primarily due to particularly noisy spectra at shorter wavelengths ($\rm{\lambda_{rf} < 2000}$ \r{A}). However, removing the high-$\chi_\nu^2$ sources does not affect the identified trends. 
Additionally, we present the $\chi_\nu^2$ measurements for two high-$z$ sources showing UV bump detections at the highest redshifts (Figs. \ref{spec}a and \ref{spec}d).

We compared the quality of fits between our fiducial model, featuring a flexible analytical dust attenuation model, and a SED fitting model using a Calzetti-like dust attenuation template for our galaxy sample\cite{2023A&A...679A..12M}. Our analysis shows that the fiducial model performs significantly better for 148 out of 173 sources ($\sim86\%$), with a $\rm{p-value} < 0.05$. The p-value distribution is heavily skewed towards lower values, emphasizing the fiducial model's superior performance. 22 sources ($\sim 13\%$) have a p-value of $0.05 - 1$ indicating that the fiducial model provides a better fit, but the improvement is not statistically significant. The fiducial model does not significantly enhance the fit for the remaining 3 sources ($\sim 2\%$), with a $\rm{p-value} = 1$. However, many sources with high p-values exhibit a flat Calzetti-like curve (constrained by the analytical attenuation model). Additionally, a few sources show poorer SED fits due to noisy spectra in the rest-frame UV. Excluding sources with high p-values from our analysis does not alter the inferred redshift trends.

Furthermore, we compared the fit to the spectra using the fiducial dust attenuation model and a bump-free model (with $c_4 = 0$)\cite{2023A&A...679A..12M}, for a subsample of 28 galaxies with a UV bump detection (right panel of Fig. \ref{z_S}). Our analysis shows that our fiducial model provides a significantly better fit for 24 out of 28 sources ($\sim 86\%$), with a $\rm{p-value} < 0.05$. This includes both high-redshift sources with UV bump detection (Fig. \ref{spec}). The p-value distribution is significantly skewed towards 0, once again emphasizing the enhanced performance of our fiducial model. For one source, a p-value $\sim 0.2$ indicates a somewhat better fit with the fiducial model. In the remaining three sources, the UV bump does not enhance the fit significantly (p-value $\sim 1$), but both models show poorer SED fits due to noisy spectra in the rest-frame UV.

\subsection*{Attenuation curve parametrization}

To compare our constrained dust attenuation curves with those documented in the literature, we employ the parameterization introduced by Salim \& Narayanan (2020)\cite{2020ARA&A..58..529S}. The primary parameters of their work that provide information about the wavelength-dependent attenuation characteristics are the UV-optical slope ($S$) and UV bump strength ($B$)\cite{2020ARA&A..58..529S}. The slope, $S$ is defined as $S  = A_{1500}/A_V$, where $A_{1500}$ is attenuation at $\lambda = 1500$ \r{A} and $A_V$ is the attenuation in the $V$-band (i.e., at $\lambda = 5500$ \r{A}). 

The UV bump strength, $B$ is defined as $B = A_{\rm{bump}}/A_{2175}$, where $A_{\rm{bump}}$ refers to the extra attenuation attributed to the bump above the baseline at $\lambda = 2175$ \r{A}, and $A_{2175}$ is the total attenuation at $\lambda = 2175$ \r{A}. We note that in the original parametrization of the UV bump with the B parameter\cite{2020ARA&A..58..529S}, the baseline at 2175 \r{A} is calculated using a relation derived from simulated attenuation curves\cite{2018ApJ...869...70N}  $A_{2175, 0} = 0.33A_{1500} + 0.67A_{A3000}$. In this work, we derive the 2175 \r{A} baseline by fitting the dust curve with an analytical dust model\cite{2008ApJ...685.1046L} and setting the $c_4$ parameter (which serves as a proxy for the UV bump strength) as $c_4 = 0$. This approach allows us to constrain the UV bump parameter $B$ more precisely, with slightly reduced uncertainties.


In Figs. \ref{all_curves} and \ref{all_curves_bin} the median attenuation curves and their dispersion are derived from the $16^{th}$, $50^{th}$, and $84^{th}$ percentiles of a set of 5000 attenuation curves that are generated using a bootstrapping approach, which involves 5000 random sampling of the $c_1-c_4$ parameters constrained for the given galaxy (sub)sample. 

\begin{figure*}
\centering
\includegraphics[width=0.49\hsize]{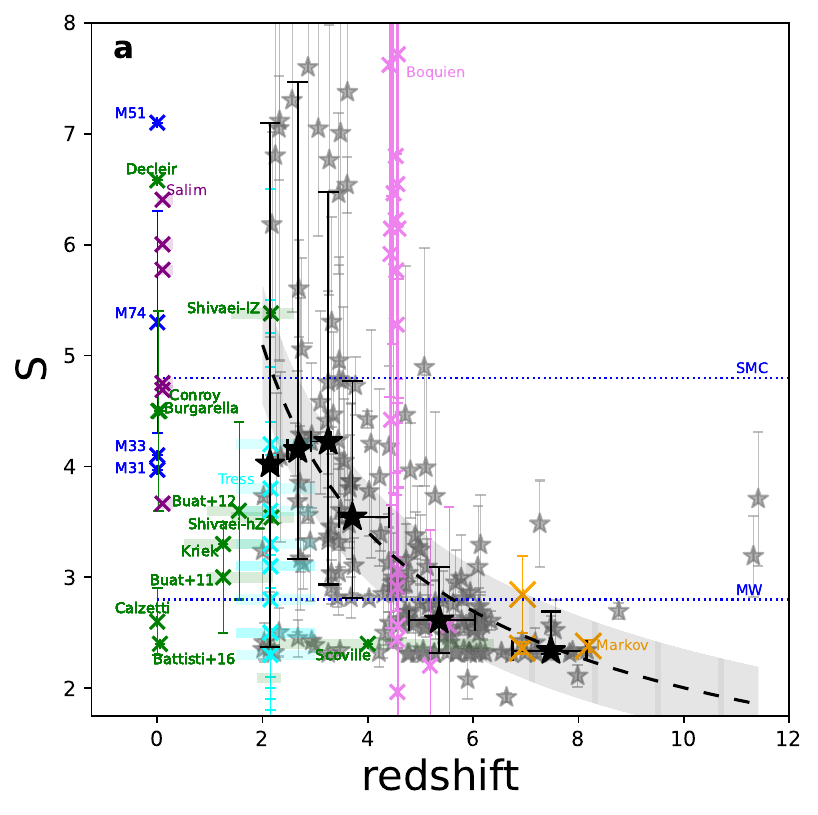}
\includegraphics[width=0.49\hsize]{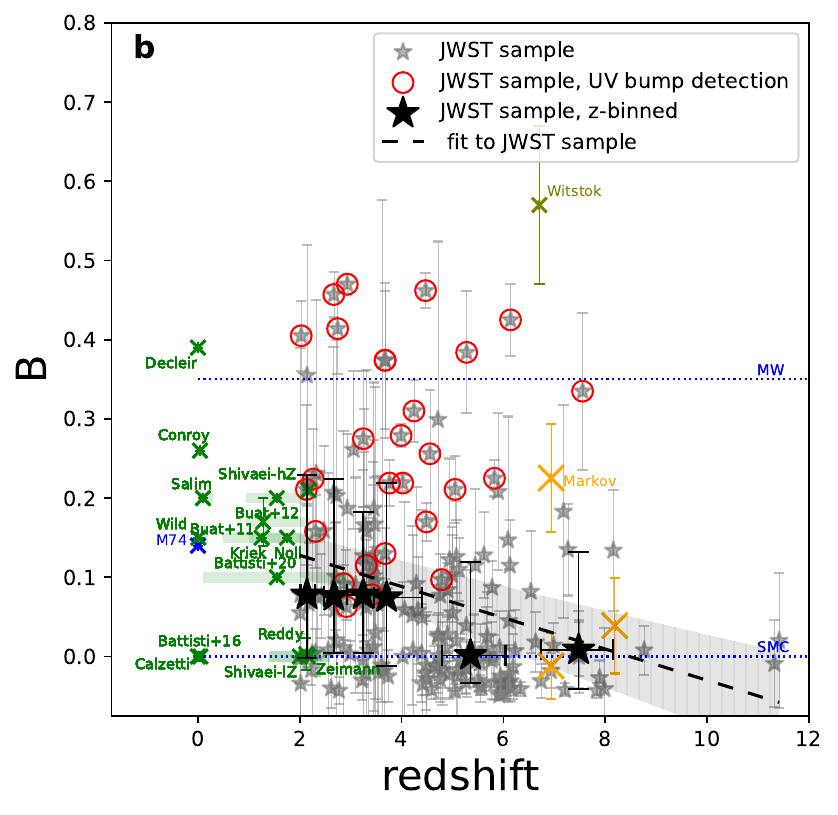}
\caption{{\bf Dust attenuation parameters as a function of redshift.} 
{\bf a}, UV-optical slope (S) as a function of redshift. {\bf b}, UV bump strength (B) as a function of redshift.  
Our sources with UV bump detection are highlighted by red circles. The slopes and the UV bumps of our galaxy sample are represented by grey stars. Corresponding error bars representing their $1\sigma$ dispersion are derived using a bootstrapping approach (See Attenuation curve parametrization section). The median slopes and UV bumps of galaxies binned by redshift (corresponding to the redshift bins of Fig. \ref{all_curves_bin}) are indicated by black stars. Corresponding errors represent their $1\sigma$ dispersion. The black dashed line and corresponding shaded region depict the best fit and $1\sigma$ uncertainty on the entire sample, derived using the \texttt{numpy polyfit} function (See Attenuation curve parametrization section). Literature results are depicted as coloured $\times$ symbols with error bars representing their $1\sigma$ uncertainties (see the Comparison with the literature section). Corresponding horizontal stripes indicate the redshift range of galaxies for which the median/mean parameters have been estimated.   
\label{z_S}
}
\end{figure*}

\begin{figure*}
\centering
\includegraphics[width=0.48\hsize]{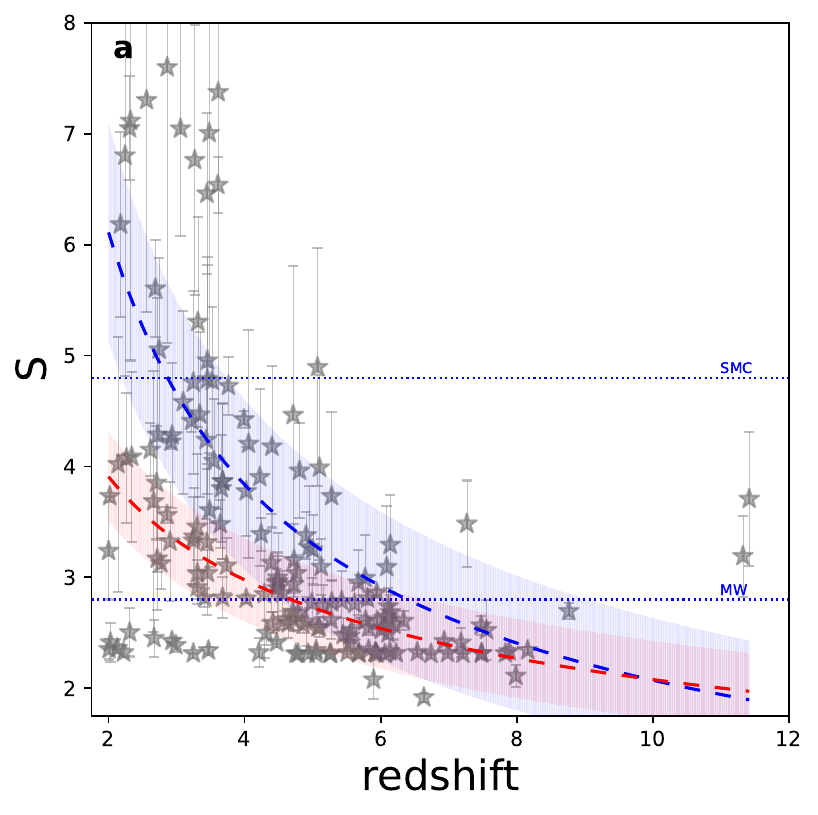}
\includegraphics[width=0.49\hsize]{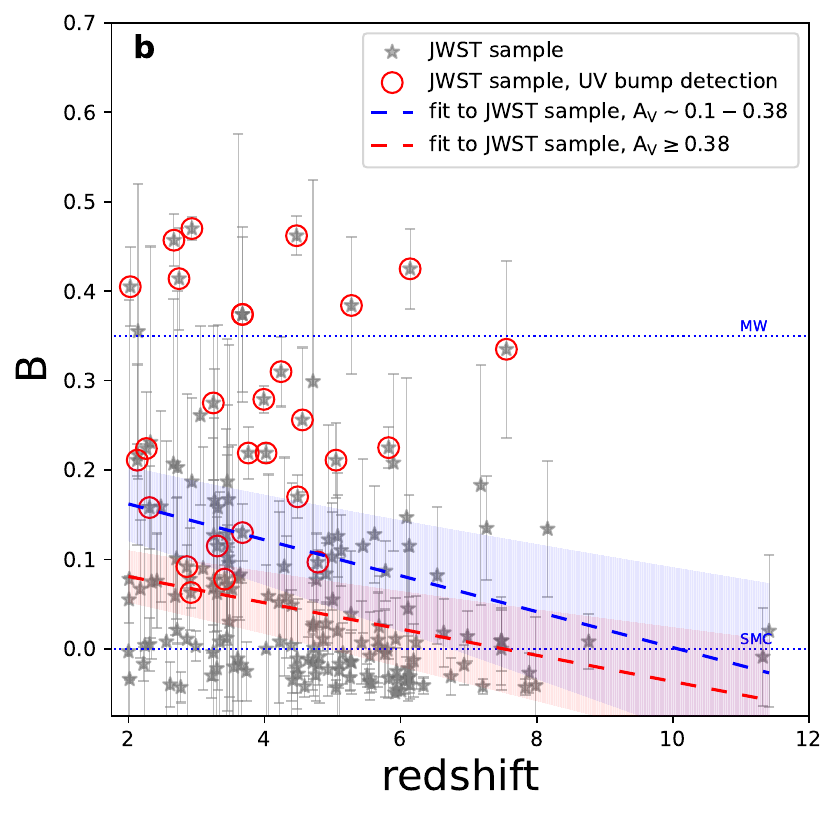}
\caption{{\bf Dust attenuation parameters as a function of redshift, binned by their $V$-band attenuation $A_V$.} 
{\bf a}, UV-optical slope (S) as a function of redshift. {\bf b}, UV bump strength (B) as a function of redshift. The slopes and the UV bumps of our galaxy sample are represented by grey stars. Corresponding error bars representing their $1\sigma$ dispersion are derived using a bootstrapping approach (See Attenuation curve parametrization section). The blue and red dashed lines and corresponding shaded regions depict the best fit and $1\sigma$ uncertainty for subsets of sources with $\rm{A_V \sim 0.1-0.38}$ (86 sources), and $\rm{A_V \geq 0.38}$ (87 sources), respectively, derived using the \texttt{numpy polyfit} function (See Attenuation curve parametrization section).
}
\label{z_S_Av}
\end{figure*}

\begin{figure*}
\centering
\includegraphics[width=0.48\hsize]{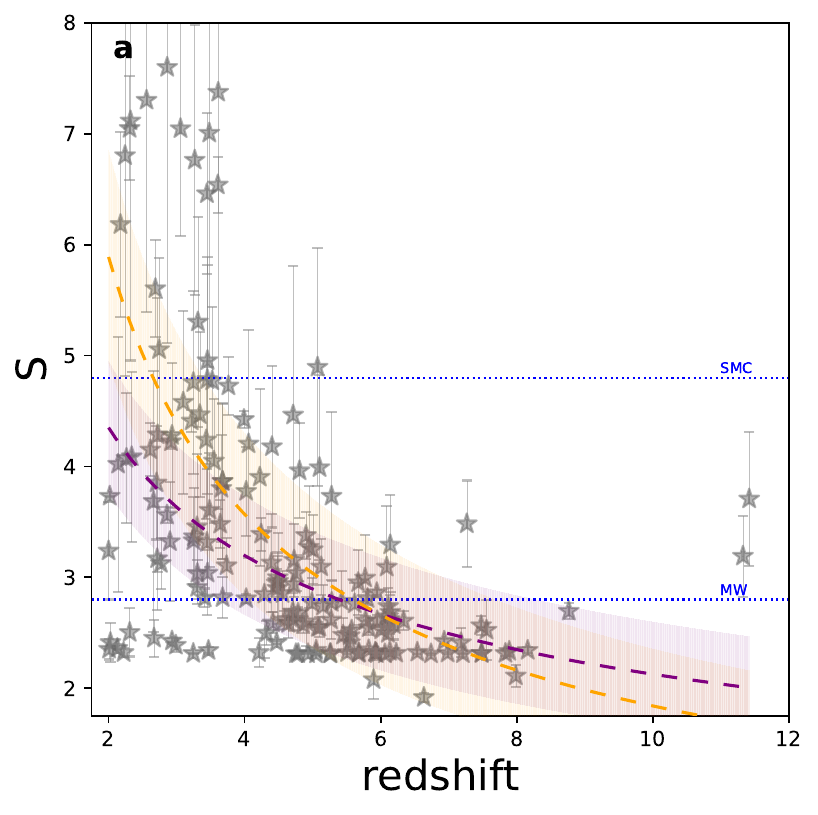}
\includegraphics[width=0.49\hsize]{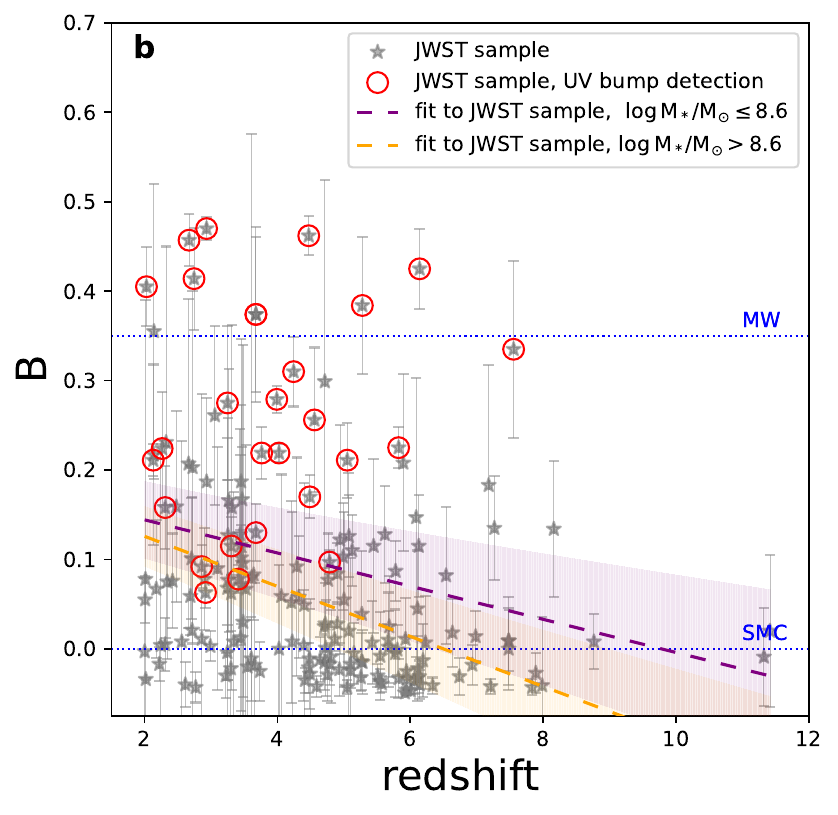}
\caption{{\bf Dust attenuation parameters as a function of redshift, binned by their stellar mass.} 
{\bf a}, UV-optical slope (S) as a function of redshift. {\bf b}, UV bump strength (B) as a function of redshift. The slopes and the UV bumps of our galaxy sample are represented by grey stars. Corresponding error bars representing their $1\sigma$ dispersion are derived using a bootstrapping approach (See Attenuation curve parametrization section). The purple and orange dashed lines and corresponding shaded regions depict the best fit and $1\sigma$ uncertainty for subsets of sources with $\rm{\log{M_*/M_{\odot}}} \leq 8.6$ (87 sources) and $\rm{\log{M_*/M_{\odot}}} > 8.6$ (86 sources), respectively, derived using the \texttt{numpy polyfit} function (See Attenuation curve parametrization section).
\label{z_S_logM}
}
\end{figure*}

\begin{figure*}
\centering
\includegraphics[width=0.48\hsize]{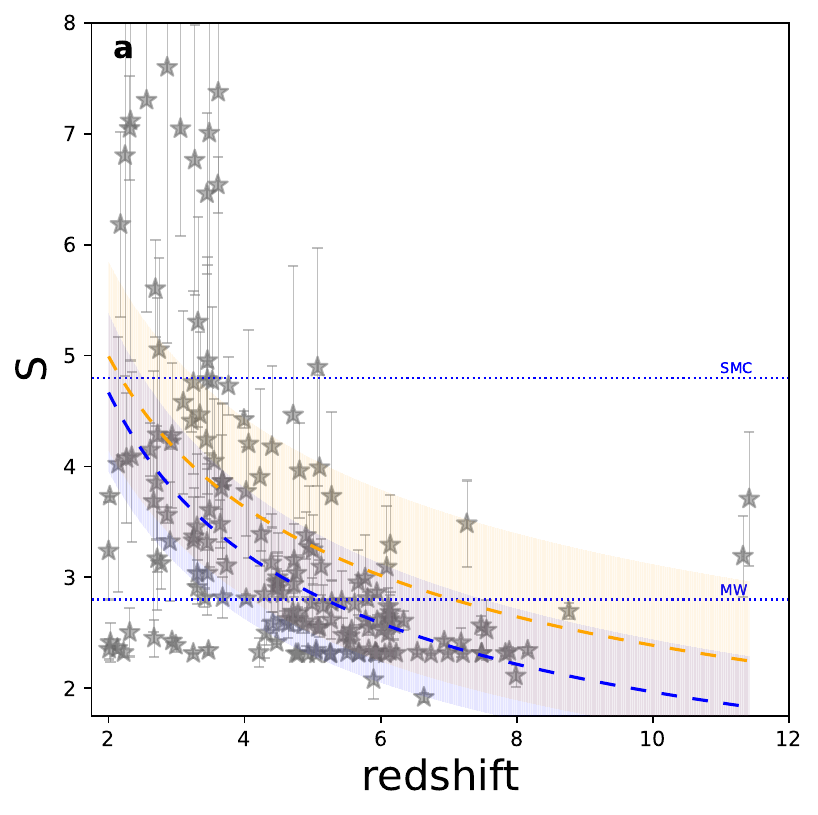}
\includegraphics[width=0.49\hsize]{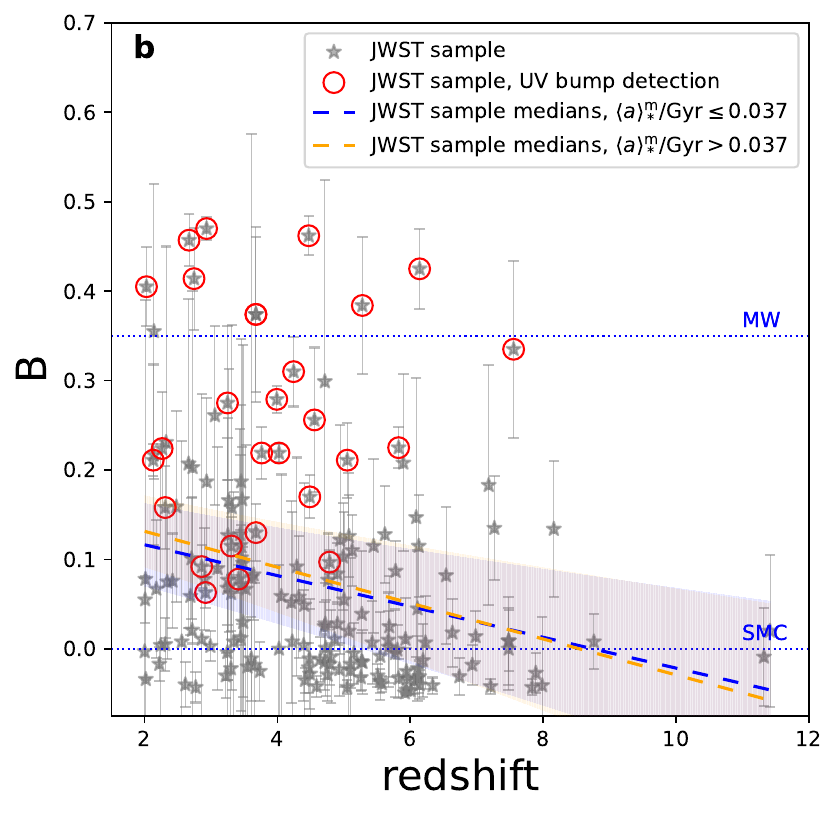}
\caption{{\bf Dust attenuation parameters as a function of redshift, binned by their mass-weighted stellar age.} 
{\bf a}, UV-optical slope (S) as a function of redshift. {\bf b}, UV bump strength (B) as a function of redshift.  The slopes and the UV bumps of our galaxy sample are represented by grey stars. Corresponding error bars representing their $1\sigma$ dispersion are derived using a bootstrapping approach (See Attenuation curve parametrization section). The blue and orange dashed lines and corresponding shaded regions depict the best fit and $1\sigma$ uncertainty for subsets of sources with ${\langle a \rangle}_*^{\rm{m}}/\rm{Gyr} \leq 0.037$ (87 sources) and ${\langle a \rangle}_*^{\rm{m}}/\rm{Gyr} > 0.037$ (86 sources), respectively, derived using the \texttt{numpy polyfit} function (See Attenuation curve parametrization section). 
\label{z_S_age}
}
\end{figure*}

\begin{figure*}
\centering
\includegraphics[width=0.48\hsize]{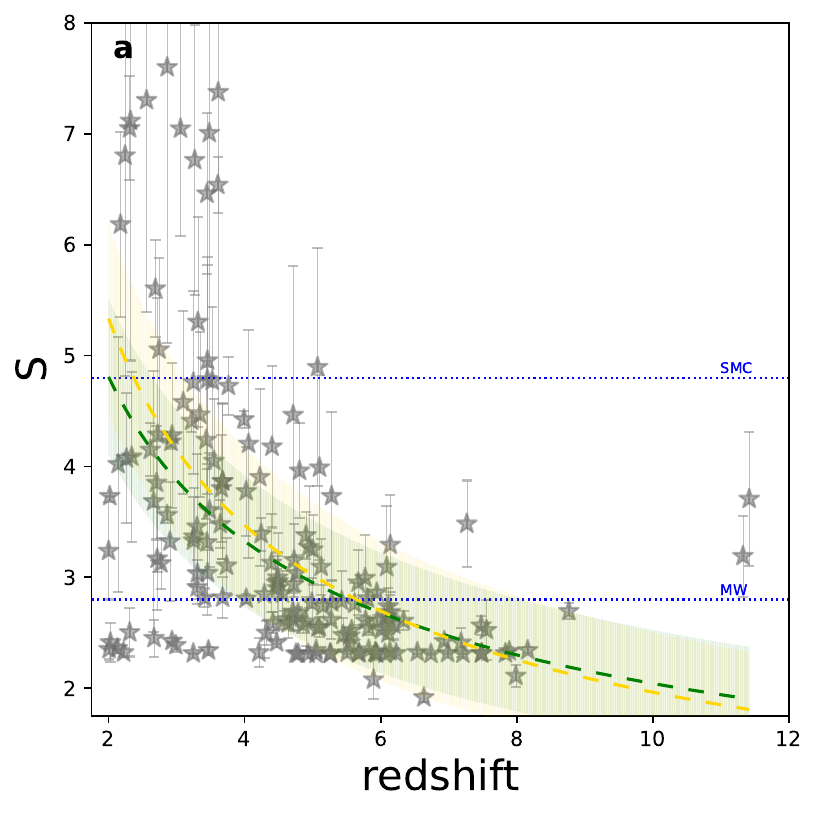}
\includegraphics[width=0.49\hsize]{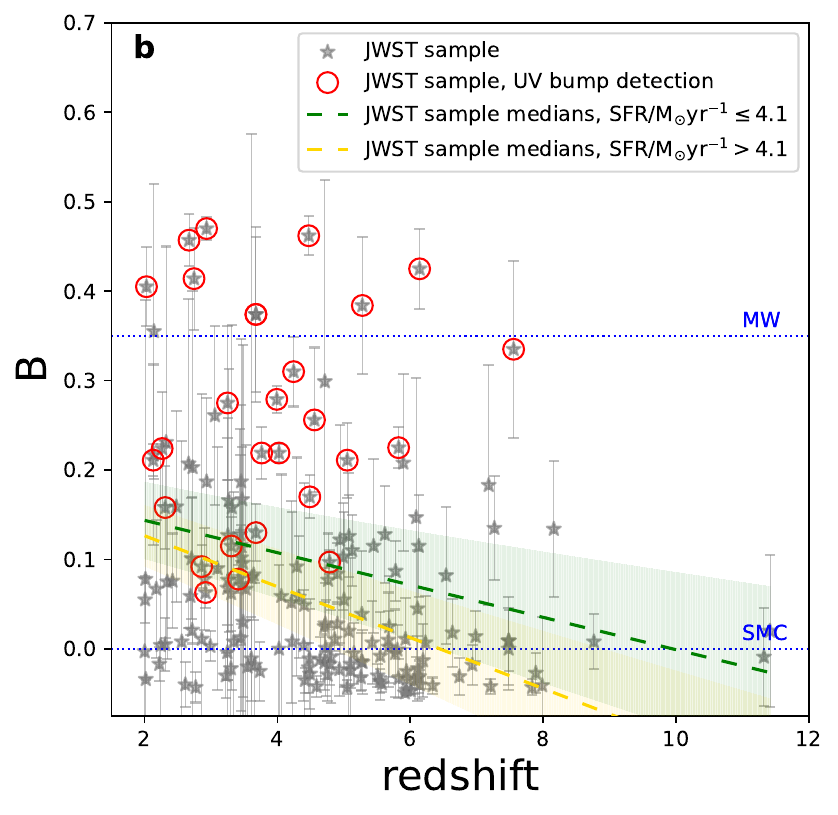}
\caption{{\bf Dust attenuation parameters as a function of redshift, binned by their SFR.} 
{\bf a}, UV-optical slope (S) as a function of redshift. {\bf b}, UV bump strength (B) as a function of redshift. The slopes and the UV bumps of our galaxy sample are represented by grey stars. Corresponding error bars representing their $1\sigma$ dispersion are derived using a bootstrapping approach (See Attenuation curve parametrization section). The green and gold dashed lines and corresponding shaded regions depict the best fit and $1\sigma$ uncertainty for subsets of sources with  $\rm{SFR}/M_{\odot} yr^{-1} \leq 4.1$ (87 sources) and $\rm{SFR}/M_{\odot} yr^{-1} > 4.1$ (86 sources), respectively, derived using the \texttt{numpy polyfit} function (See Attenuation curve parametrization section). 
\label{z_S_SFR}
}
\end{figure*}

\begin{figure*}
\centering
\includegraphics[width=0.48\hsize]{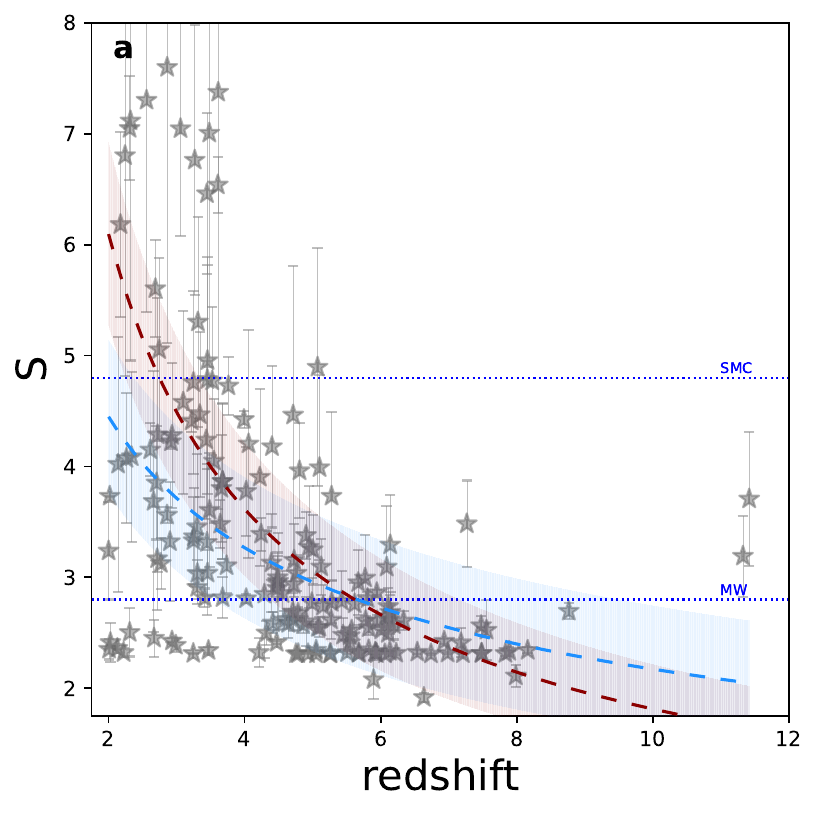}
\includegraphics[width=0.49\hsize]{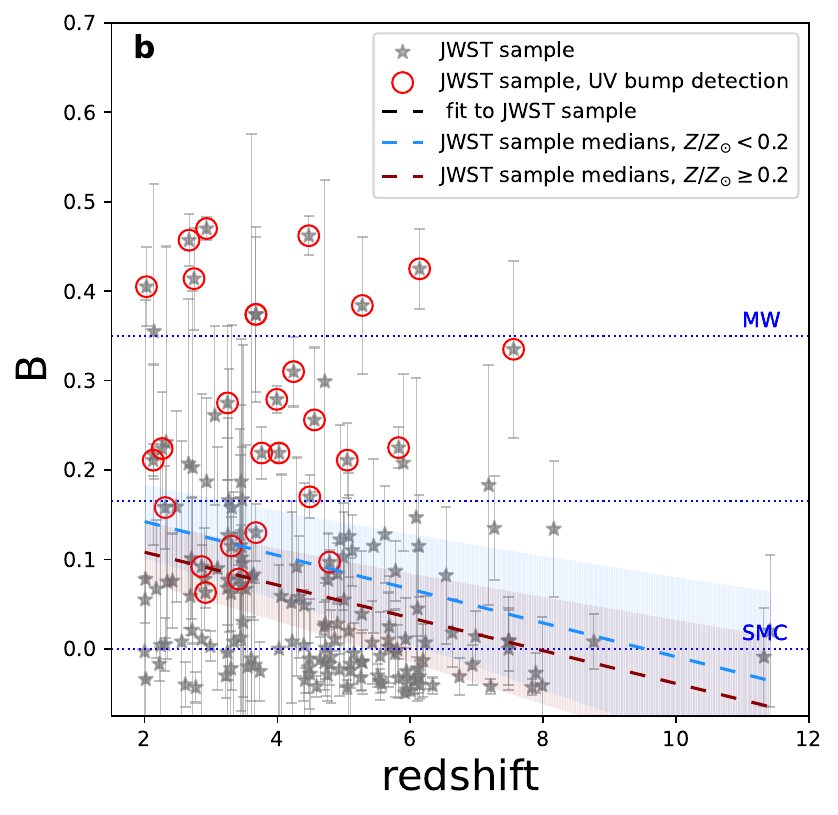}
\caption{{\bf Dust attenuation parameters as a function of redshift, binned by their metallicity.} {\bf a}, UV-optical slope (S) as a function of redshift. {\bf b}, UV bump strength (B) as a function of redshift. The slopes and the UV bumps of our galaxy sample are represented by grey stars. Corresponding error bars representing their $1\sigma$ dispersion are derived using a bootstrapping approach (See Attenuation curve parametrization section). The light blue and dark red dashed lines and corresponding shaded regions depict the best fit and $1\sigma$ uncertainty for subsets of sources with metallicities of $Z/Z_{\odot} < 0.199$ (85 sources) and $Z/Z_{\odot} \geq 0.199$ (88 sources), respectively, derived using the \texttt{numpy polyfit} function (See Attenuation curve parametrization section). 
\label{z_S_Z}
}
\end{figure*}

\begin{figure*}
\centering
\includegraphics[width=0.48\hsize]{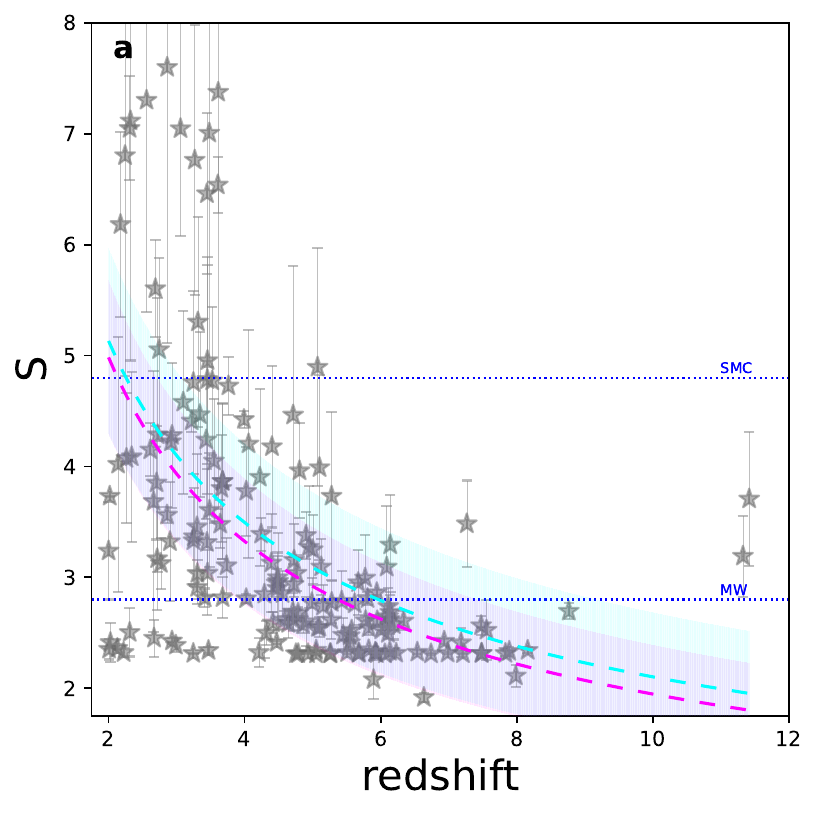}
\includegraphics[width=0.49\hsize]{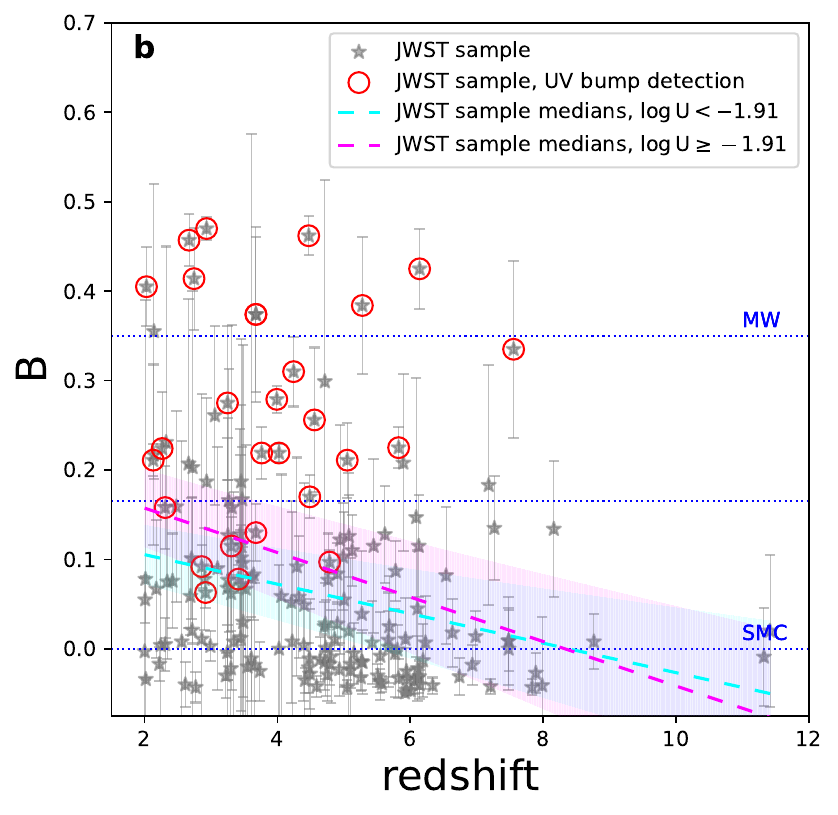}
\caption{{\bf Dust attenuation parameters as a function of redshift, binned by their ionization parameter.}  {\bf a}, UV-optical slope (S) as a function of redshift. {\bf b}, UV bump strength (B) as a function of redshift. The slopes and the UV bumps of our galaxy sample are represented by grey stars. Corresponding error bars representing their $1\sigma$ dispersion are derived using a bootstrapping approach (See Attenuation curve parametrization section). The cyan and magenta dashed lines and corresponding shaded regions depict the best fit and $1\sigma$ uncertainty for subsets of sources with $\rm{\log{U}} < -1.91$ (87 sources) and $\rm{\log{U}} \geq -1.91$ (86 sources), respectively, derived using the \texttt{numpy polyfit} function (See Attenuation curve parametrization section). 
\label{z_S_logU}
}
\end{figure*}

\begin{figure}
\centering
\includegraphics[width=\hsize]{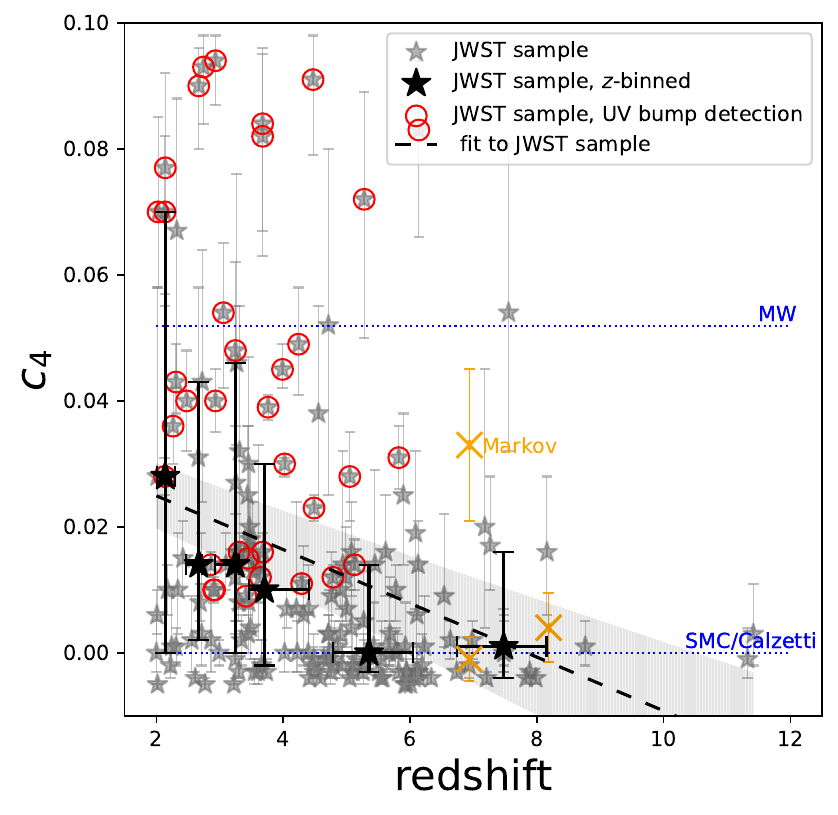}
\caption{{\bf UV bump strength, parametrized by the $c_4$ parameter\cite{2008ApJ...685.1046L}, as a function of redshift.}  
Our entire galaxy sample is represented by grey stars. The UV bump strengths and their associated $1\sigma$ uncertainties are derived from the medians and $1\sigma$ dispersions of the $c_4$ parameter's posterior distribution (See Attenuation curve parametrization section). Sources with UV bump detection are highlighted by red circles. The median UV bumps of galaxies binned by redshift (corresponding to the redshift bins of Fig. \ref{all_curves_bin}) are indicated by black stars. Corresponding errors represent their $1\sigma$ dispersion. The black dashed line and corresponding shaded region depict the best fit and $1\sigma$ uncertainty on the entire sample, derived using the \texttt{numpy polyfit} function (See Attenuation curve parametrization section). Literature results are depicted as orange $\times$ symbols with error bars representing their $1\sigma$ uncertainties (see the Comparison with the literature section).
\label{z_c4}
}
\end{figure}

\begin{figure*}
\centering
\includegraphics[width=0.56\hsize]{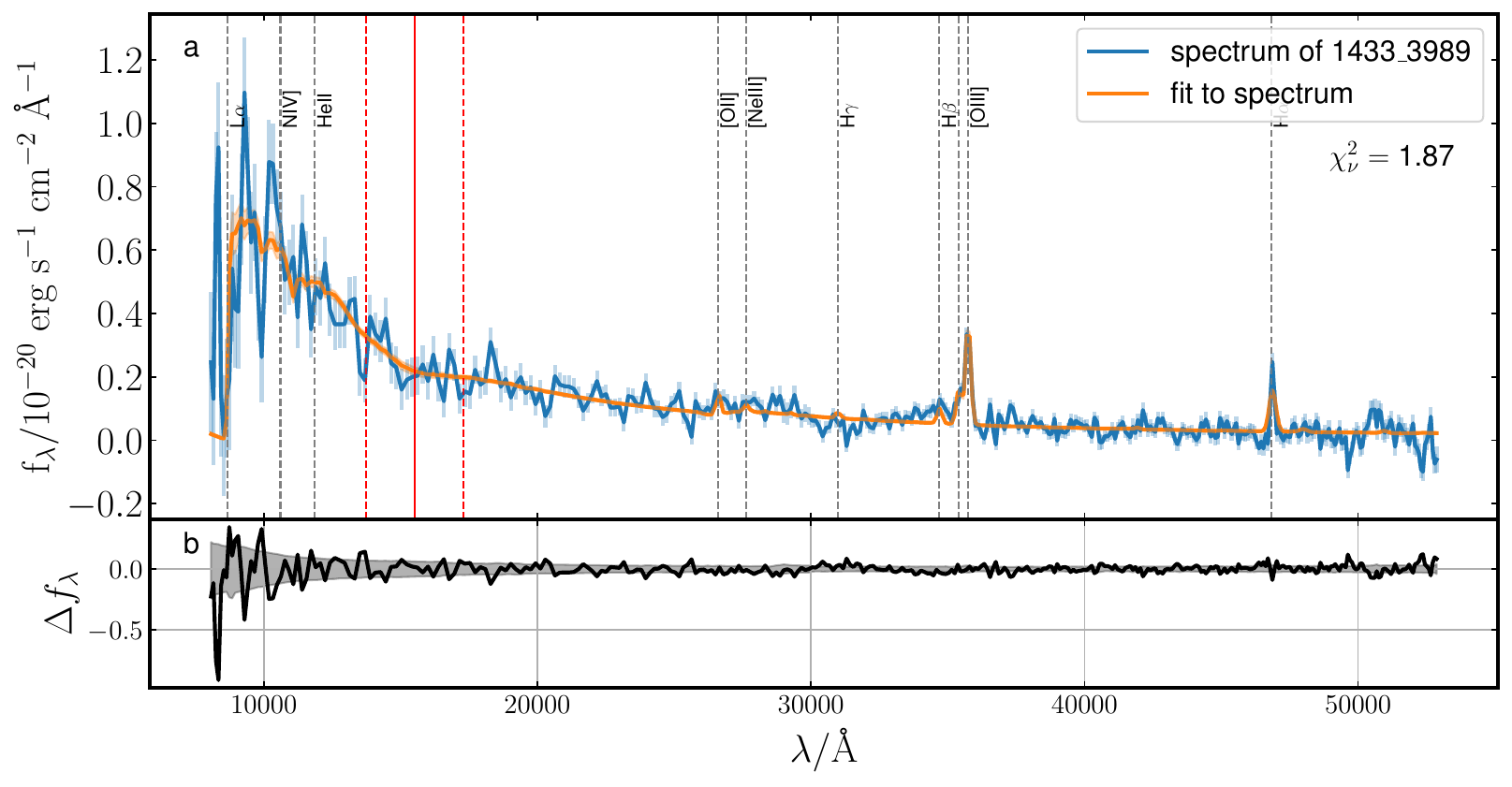}
\includegraphics[width=0.38\hsize]{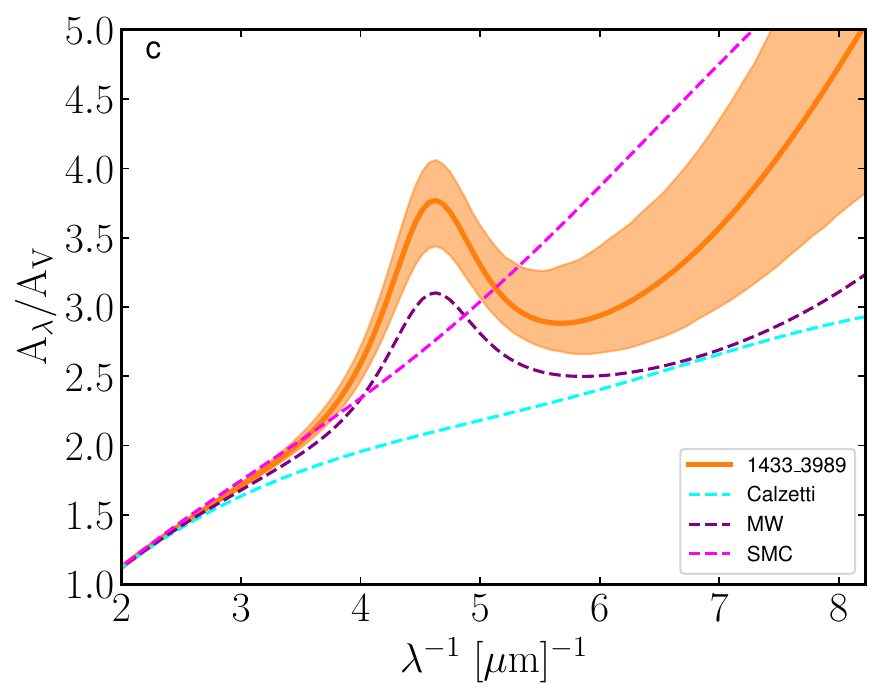}
\includegraphics[width=0.56\hsize]{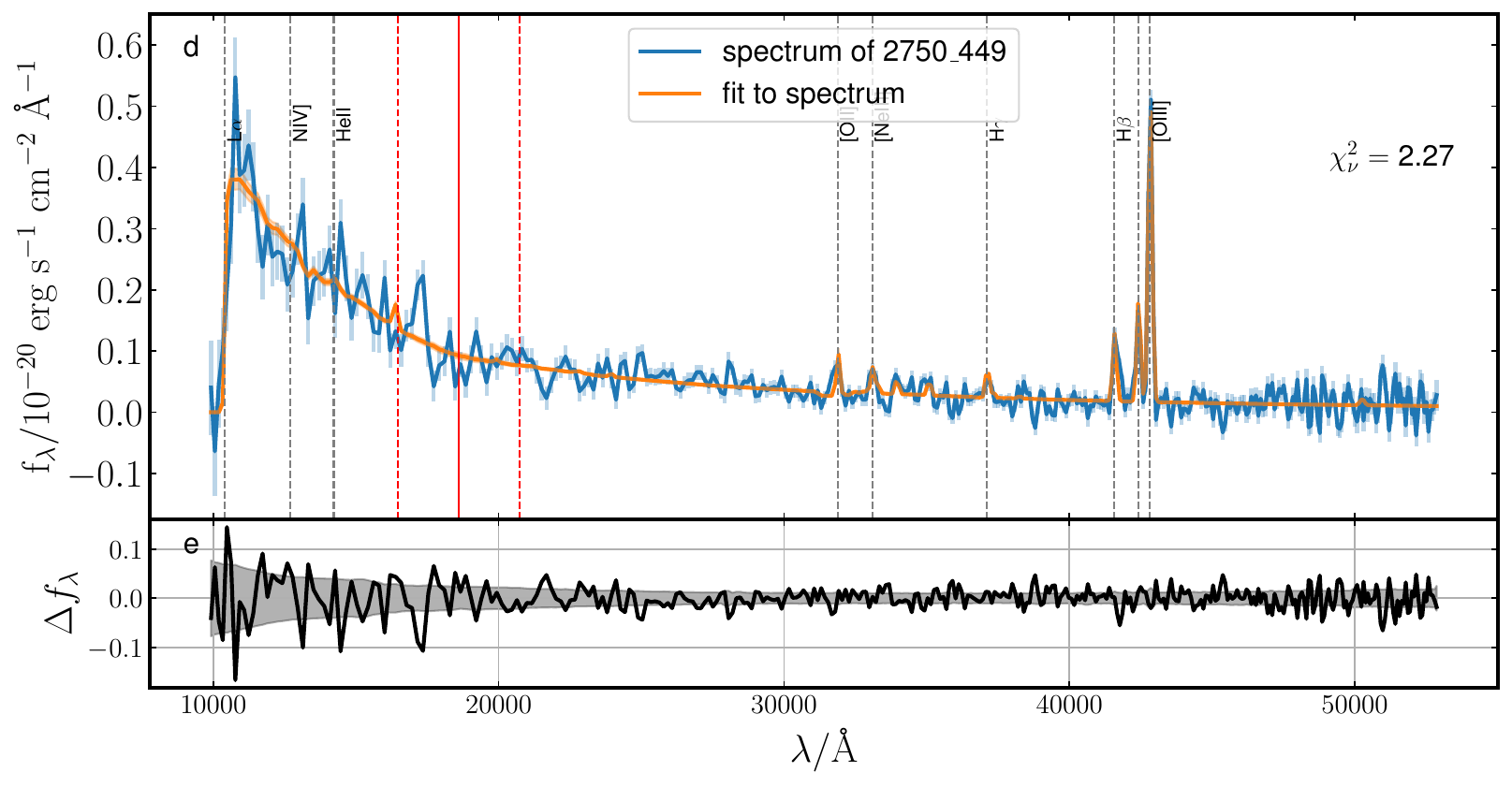}
\includegraphics[width=0.38\hsize]{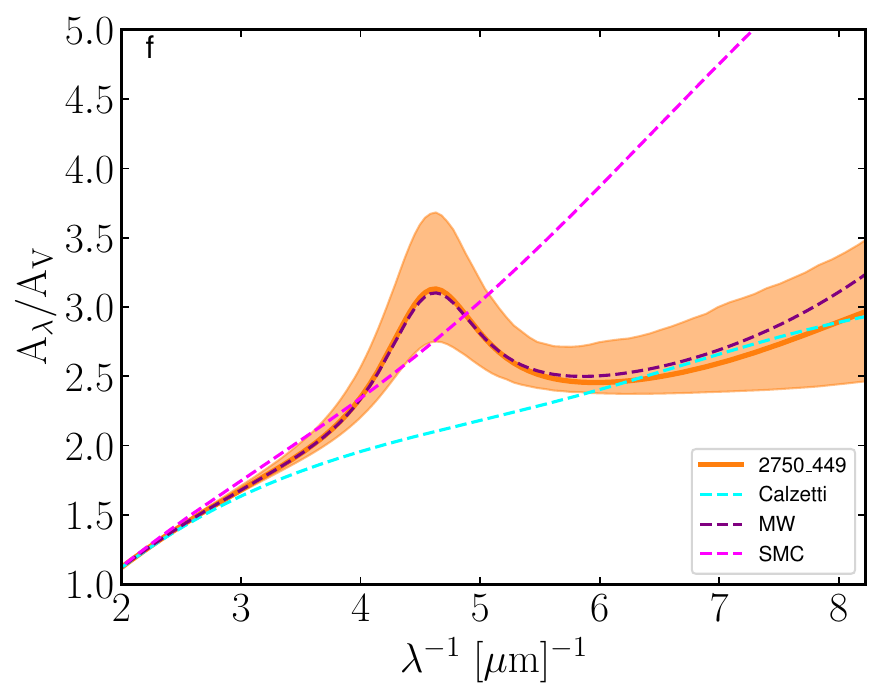}
 \caption{{\bf SED fitting results for two high-redshift sources exhibiting UV-bump detection.} Top row (panels {\bf a - c}): SED fit for the $1433\_3989$ galaxy at $z \approx 6.14$. Bottom row (panels d - f): SED fit for $2750\_449$ at $z \approx 7.55$. {\bf a, d}, The NIRSpec {\it JWST} spectrum and the best-fit posterior, along with their corresponding $1\sigma$ uncertainties are indicated in blue and orange, respectively. Vertical grey and red lines mark the positions of potential emission lines and the UV bump absorption feature of the spectrum. {\bf b, e},
 Residuals of the best fit on the observed spectrum, $\Delta f_{\lambda}$, with 1$\sigma$ uncertainties. {\bf c, f}, The  best-fit dust attenuation curve with $1\sigma$ uncertainties.  The uncertainties are estimated using a boot-strap method involving generating 5000 attenuation curves from
a random sampling of the $c_1 - c_4$ from the posterior. Dust model fits to the Calzetti, the MW, and the SMC empirical curves are depicted as cyan, purple, and magenta dashed lines, respectively.
 }
 \label{spec}
\end{figure*}

In Figs. \ref{z_S} $-$ \ref{z_S_Av}  and Extended Data Figs. \ref{z_S_logM} $-$ \ref{z_S_logU} the values of $S$ and $B$ for each galaxy (grey stars) are determined through the application of the dust attenuation model, $A_{\lambda}/A_V = f(\lambda, c_1, c_2, c_3, c_4)$ (Eq. \ref{dust_law}), where the $c_1-c_4$ parameters are the outputs of the SED fitting procedure for each source. Next, the uncertainty on $S$ and $B$ represents $1\sigma$ dispersion from the median, estimated by employing a boot-strapping approach, which involves generating 100 attenuation curves $A_{\lambda}/A_V$ from a random sampling of the $c_1-c_4$ parameters from the posterior. In Fig. \ref{z_S}, the median values and 1$\sigma$ dispersion of $S$ and $B$ parameters of subsets of galaxies grouped by redshift (black stars) are estimated directly from the distribution of these parameters within each subset.

In Figs. \ref{z_S} $-$ \ref{z_S_Av} and Extended Data Figs. \ref{z_S_logM}$-$\ref{z_S_logU},  the slope-redshift and UV bump-redshift relations have the form $\log{S} = k_1 \log{z} + n_1$ and and $B = k_2 z + n_2$. The best-fit values for the model parameters $k_i$  and $n_i$, along with their $1\sigma$ uncertainties are obtained using a \texttt{numpy polyfit} function, as

\begin{equation}
    \log{S} = (-0.580 \pm 0.064) \log{z} + (0.882 \pm 0.042),
\end{equation}

\begin{equation}
    B = (-0.020 \pm 0.05) z + (0.167 \pm 0.025).
\end{equation}

The uncertainties on the best-fit values of $S$ and $B$, are calculated by propagating the errors in the fitted parameters $k_i$  and $n_i$, and represents the $1\sigma$ standard error of the predicted values of $S$ and $B$.

In Extended Data Fig. \ref{z_c4}, the $c_4$ parameter along with its associated uncertainty for each source represents the median and 1$\sigma$ dispersion of the posterior distribution. The values and uncertainties of the $c_4$ parameter for a subset of galaxies grouped by redshift, as well as the $c_4$-redshift relation are determined analogously as the $B$ parameter in Fig. \ref{z_S}b.

We have compared the V-band attenuation, $A_V$, derived from SED fitting to values from the Balmer decrement, $A_V^{\rm{BD}}$. We find that over $25\%$ of the galaxies in the sample are characterised by negative $A_V^{\rm{BD}}$\cite{2015ApJ...806..259R, 2023ApJ...949L..11M, 2023arXiv230603931S} values. This non-physical result arises from observed line ratios that are lower or comparable to the intrinsic value of $2.86-2.87$\cite{2006agna.book.....O}.

The Balmer line ratio can be affected by various issues, including low spectral resolution, poor line fits, assumptions on empirical relations (e.g., H$\alpha$/NII line ratio), and physical conditions (e.g., temperature and gas density)\cite{2023MNRAS.518..425C, 2023ApJ...952..167R} at high-$z$ that differ from the standard ones\cite{2006agna.book.....O}. Therefore, for consistency and reliability, we have used $A_V$ estimates obtained from SED fitting. 

\subsection*{Comparison with the literature}

In Fig. \ref{z_S} we report the literature findings for the slope $S$ and the bump $B$. The MW\cite{1989ApJ...345..245C} and the SMC\cite{2003ApJ...594..279G} are indicated by blue dotted lines.  Other local individual sources: M31\cite{2017A&A...599A..64V}, M33\cite{2019MNRAS.487.2753W}, M51\cite{2014A&A...571A..69D}, and M74\cite{2019MNRAS.486..743D} are depicted as blue $\times$ symbols. The median/mean values of samples of nearby ($z \sim 0$) and intermediate-$z$ ($z \sim 1-4$) galaxies from various studies\cite{2000ApJ...533..682C, 2005MNRAS.360.1413B, 2010ApJ...718..184C, 2011MNRAS.417.1760W, 2011A&A...533A..93B, 2012A&A...545A.141B, 2013ApJ...775L..16K, 2015ApJ...806..259R, 2015ApJ...800..108S, 2016ApJ...818...13B, 2019PhDT.......129D, 2020ApJ...899..117S} are represented by green $\times$ symbols. Slopes for subsamples of low and intermediate-$z$ galaxies are shown as magenta\cite{2018ApJ...859...11S} and cyan\cite{2018MNRAS.475.2363T} $\times$ symbols, respectively. Finally, individual high-$z$ ($z \sim 4-8$) sources are indicated by violet\cite{2022A&A...663A..50B}, olive\cite{2023Natur.621..267W} and orange $\times$ symbols\cite{2023A&A...679A..12M}.

Literature results for all the local sources and most nearby\cite{2000ApJ...533..682C, 2005MNRAS.360.1413B, 2010ApJ...718..184C, 2011MNRAS.417.1760W, 2016ApJ...818...13B} and intermediate-$z$\cite{2009A&A...499...69N, 2011A&A...533A..93B, 2012A&A...545A.141B, 2013ApJ...775L..16K, 2015ApJ...806..259R, 2015ApJ...800..108S} galaxy samples are directly adopted from the work of Salim \& Narayanan (2020)\cite{2020ARA&A..58..529S}. Specific literature values of $S$ and $B$ are recalculated from their original parametrizations using the relations provided by Salim \& Narayanan (2020)\cite{2020ARA&A..58..529S}. For instance, certain studies\cite{2018ApJ...859...11S} utilize the Charot \& Fall (2000) parametrization\cite{2000ApJ...539..718C}, $A_{\lambda}/A_V  = (\lambda/0.55 \mu m)^{-n}$, where the power law exponent $n$ characterizes the UV-optical slope. In such cases, we make use of the relation $\log{S} = n/1.772$ to convert to the $S$ parameter. In works that use the Noll et al. (2009) parametrization\cite{2009A&A...499...69N, 2019PhDT.......129D}, and the $\delta$ parameter to parametrize the slope, we use the relation $\log{S} = 0.4 - 0.55\delta$. Tress et al. (2018)\cite{2018MNRAS.475.2363T} use the Conroy et al. (2010) parametrization\cite{2010ApJ...718..184C} and they provide a relation to convert to the Noll et al. (2009) parametrization\cite{2009A&A...507.1793N}. Finally, the remaining literature results\cite{2019PhDT.......129D, 2020ApJ...899..117S, 2023Natur.621..267W, 2023A&A...679A..12M} are directly provided by the authors.

To date, the detection of the characteristic 2175 \r{A} feature has been reported in two galaxies at $z\sim 7$\cite{2023Natur.621..267W, 2023A&A...679A..12M}, both of which are also a part of the DAWN sample. 
Witstok et al. (2023)\cite{2023Natur.621..267W} reported a UV bump detection (with a $6.4\sigma$) in the JADES-GS-z6-0 galaxy, achieved by fitting an excess attenuation centred at 2236 \r{A} with a Drude profile. However, applying our method (but setting the central wavelength of the bump to 2236 \r{A}, in order to match the peak wavelength of the UV bump reported in their study) to the full spectrum of JADES-GS-z6-0 from JADES resulted in a marginal detection, with a UV bump strength of $B = 0.134 \pm 0.067$ or $c_4 = 0.020_{-0.011}^{+0.012}$ ($\sim 2\sigma$). However, our model assumes a UV bump width that is $\sim$ twice as large as the value used in Witstok et al. (2023). Differences in the assumed bump width between the two studies could explain the variation in results.

Next, in our recent study\cite{2023A&A...679A..12M}, we reported a hint of a UV bump detection in one out of three high-$z$ sources, s00717 at $z \sim 6.9$, with a UV bump strength of $c_4 = 0.033_{-0.012}^{+0.013}$ ($\sim 2.7\sigma$). Employing the identical SED fitting method with the same set of priors\cite{2023A&A...679A..12M} on the DAWN observations of the s00717 source yields $c_4 = 0.019_{-0.017}^{+0.031}$.

Inconsistencies in these results may arise from the use of various fitting procedures involving different models (for JADES-GS-z6-0) as well as different data reduction pipelines generating slightly different final 1D spectra (for s00717). For s00717, the data reduction in our previous work was performed with the standard pipeline\cite{2022A&A...661A..80J}, whereas the data in this work are reduced with the \texttt{MSAEXP} custom code\cite{Brammer_msaexp_NIRSpec_analyis_2022}.
The different pipeline versions have different sampling, resulting in slightly different flux densities, uncertainties, and channel widths, which can create inconsistencies. When investigating the redshift evolution of galaxy properties, it is thus recommended to adopt samples of galaxies for which the data reduction analysis is performed with the same pipeline, as in the case of our study. 

\subsection*{Observational and model uncertainties}

Factors that could contribute to the discrepancy in slopes between our study and the ones from the literature at $z \sim 2$ can arise both from observational uncertainties and model dependencies. Regarding the observational uncertainties, at the low redshift limit of our sample ($z \sim 2-2.8$), NIRSpec prism observations ($\lambda \sim 0.6-5.3 \ \mu m$) are restricted to the rest-frame $\lambda  \gtrsim 1600-2000$ \r{A}, which probe the UV bump peak and marginally extend into the far-UV rise of the attenuation curve. This constraint in wavelength coverage may contribute to larger uncertainties in inferred slopes (Fig. \ref{all_curves_bin}). Moreover, for $z \sim 2-2.8$ sources, NIRSpec data are probing the attenuation curve with a fraction of the spectra at shorter wavelengths, which suffer from larger uncertainties,  due to the lower resolution ($R \sim \lambda/\Delta \lambda \sim 50$) of the NIRSpec prism disperser in the wavelength range of $\lambda_{\rm{obs}} \sim 0.5-2 \ \mu m$. Finally, the observed discrepancy may be attributed to the lower number of sources per bin at intermediate redshifts (13-21 sources) compared to the higher redshift bins (20-66 objects).

Regarding the model dependencies, different SED fitting codes with different sets of model prescriptions (SFHs\cite{2023A&A...679A..12M, 2023MNRAS.519.5859W, 2023MNRAS.522.6236T}, dust attenuation, SPS and initial mass function (IMF)\cite{2022MNRAS.510.5603K, 2024ApJ...963...74W}) and imposed priors on parameters\cite{2012A&A...545A.141B, 2018MNRAS.475.2363T} can have a significant impact on the output parameters, including the recovered attenuation curve.  We conducted extensive testing of our customized SED fitting tool with seven different SFH models\cite{2023A&A...679A..12M}. Our analysis revealed that while the choice of SFH model can influence properties closely linked to the SFH itself (e.g., $M*$, SFR, and ${\langle a \rangle}_*^{\rm{m}}$), the constrained dust attenuation properties remain largely unaffected by the assumed SFH. This finding underscores the robustness of our results concerning dust attenuation, irrespective of the specific SFH model employed.

\section*{Data availability}

The data that support the findings of this study are available at  \url{https://dawn-cph.github.io/dja/index.html}.

\section*{Code availability}

 \texttt{GRIZLI} (Grism redshift and line analysis software for space-based slitless spectroscopy\cite{brammer_2023_8370018}) and \texttt{MSAEXP} (Tool for extracting JWST NIRSpec) Micro-Shutter Assembly (MSA) spectra\cite{Brammer_msaexp_NIRSpec_analyis_2022}, and \texttt{BAGPIPES} (Bayesian Analysis of Galaxies for Physical Inference and Parameter EStimation)\cite{2018MNRAS.480.4379C} Python scripts are publicly available at \url{https://github.com/gbrammer/grizli}, \url{https://github.com/gbrammer/msaexp}, and  \url{https://github.com/ACCarnall/bagpipes}, respectively.


\section*{Acknowledgements}

We would like to express our gratitude to H. Hirashita and I. Shivaei for engaging in productive discussions with us and providing valuable insights and comments. Additionally, we thank M. Decleir, I. Shivaei, and J. Witstok for sharing their findings on the properties of the attenuation curve. 
       VM, AF, AP, and MK acknowledge support from the ERC Advanced Grant INTERSTELLAR H2020/740120. VM acknowledges support from the ERC Grant FIRSTLIGHT and from the Slovenian National Research agency
       ARRS through grants N1-0238, P1-0188. Partial support (AF) from the Carl Friedrich von Siemens-Forschungspreis der Alexander von Humboldt-Stiftung Research Award is kindly acknowledged.
       Any dissemination of results must indicate that it reflects only the author’s view and that the Commission is not responsible for any use that may be made of the information it contains.
       The data products presented herein were retrieved from the Dawn JWST Archive (DJA). DJA is an initiative of the Cosmic Dawn Center, which is funded by the Danish National Research Foundation under grant No. 140.
       We gratefully acknowledge the computational resources of the Center for High Performance Computing (CHPC) at SNS.

\section*{Author contributions}

VM, SG, and AF led the writing of this paper. EP, VM, MK, and SG contributed to the target selection. VM, EP, and SG contributed to the visual inspection and additional data reduction. SG, VM, and AP contributed to the sample cleaning. VM led the SED fitting of the selected targets. SG, VM, AF, AP, LS, and FDM contributed to the development of the customized SED fitting tool. VM, SG, LS, AP, AF, and MK contributed to the visualisation of the results. VM, SG, AP, MK, AF, EP, and FDM contributed to the analysis of the results. VM, AF, SG, and AP contributed to the discussion on the physical interpretation of the results. All authors reviewed the manuscript. 

\section*{Competing interests}  The authors declare no competing interests.

\bibliography{main}

\section*{Extended data}

\newpage

\end{document}